\documentclass[aps,pra,manuscript]{revtex4}
\usepackage{graphicx}
\usepackage{epsfig}
\usepackage{color}
\usepackage{bm}
\begin{document}
\draft
\title{Controlling quantum-dot light absorption and emission by a surface-plasmon field}

\author{Danhong Huang$^{1}$, Michelle Easter$^{2}$, Godfrey Gumbs$^{3}$,\\
A. A. Maradudin$^{4}$, Shawn-Yu Lin$^{5}$, Dave Cardimona$^{1}$,\\
and Xiang Zhang$^{6}$}
\affiliation{$^{1}$Air Force Research Laboratory, Space Vehicles
Directorate, Kirtland Air Force Base, New Mexico 87117, USA\\
$^{2}$Department of Mechanical Engineering, Stevens Institute of Technology, 1 Castle Point Terrace, Hoboken, New Jersey 07030, USA\\
$^{3}$Department of Physics and Astronomy, Hunter College of the City University of New York, 695 Park Avenue New York, New York 10065, USA\\
$^{4}$Department of Physics and Astronomy,
University of California, Irvine, California 92697, USA\\
$^{5}$Department of Electrical, Computer and Systems Engineering,\\
Rensselaer Polytechnic Institute, 110 8th Street, Troy, New York 12180, USA\\
$^{6}$Department of Mechanical Engineering, 3112 Etcheverry Hall,\\
University of California at Berkeley, Berkeley, California 94720}

\date{\today}

\begin{abstract}
The possibility for controlling the probe-field optical gain and absorption switching and photon conversion by a surface-plasmon-polariton near field is explored for a quantum dot above the surface of a metal.
In contrast to the linear response in the weak-coupling regime, the calculated spectra show an induced optical gain and a triply-split spontaneous emission peak resulting from the interference between the surface-plasmon field
and the probe or self-emitted light field in such a strongly-coupled nonlinear system. Our result on the control of the mediated photon-photon interaction,
very similar to the `gate' control in an optical transistor, may be experimentally observable and applied to
ultra-fast intrachip/interchip optical interconnects, improvement in the performance of fiber-optic communication networks and developments of optical digital computers and quantum communications.
\end{abstract}
\pacs{PACS:}
\maketitle

\section{Introduction}
\label{sec1}

It is well known that photons inherently do not interact with each other.
In classical electrodynamics, the Maxwell equations are linear and cannot describe any photon-photon interaction. However, effective photon-photon coupling could exist in a mediated way, e.g. through their
direct interactions with matter. Very recently, an experiment\,\cite{lukin}, which involves firing pairs of photons through an ultra-cold atomic gas, was reported to provide the evidence for an attractive
interaction between the photons to form the so called `molecules' of light. In general, if the interaction between photons and matter is strong, the optical response of matter will become nonlinear and
the resulting bandedge optical nonlinearities\,\cite{r7} will enable an effective photon-photon interaction.\,\cite{sci}
An optical transistor\,\cite{otrans} could be built based on this basic idea, where `gate' photons control the intensity of a `source' light beam.
Optical transistors could be applied to speed up and improve the performance of fiber-optic communication networks. Here, all-optical digital signal processing and routing is fulfilled by arranging optical transistors
in photonic integrated circuits and the signal loss during the propagation could be compensated by inserting new types of optical amplifiers. Moreover, optical transistors are expected to play an important role
in the developments of an optical digital computer or quantum-encrypted communication.
\medskip

Most previous research on optical properties of materials, including optical absorption, inelastic light scattering and spontaneous emission, used a weak probe field as a perturbation to the studied system.\,\cite{book}
In this weak-coupling regime, the optical response of electrons depends only on the material characteristics,\,\cite{linear} and therefore, no photon-photon interaction is expected. However, the strong-coupling regime
could be reached with help from microcavities and the experimental effort on
searching for polariton condensation (resulting from strong light-electron interaction) in semiconductors continues to produce results.\,\cite{polar1,polar2,polar3}
The general review of exciton-polariton condensation can be found from Ref.\,[\onlinecite{rmp}].
The successful demonstration of room-temperature polariton lasing without population inversion in semiconductor microcavities using both optical pumping\,\cite{pump1,pump2} and electrical injection\,\cite{injec1,injec2}
have made it possible for ultra-low lasing thresholds and very-small emitter sizes comparable to the emitted wavelength.
Semiconductor exciton-polariton nanolasers could advance intrachip and interchip optical interconnects by integrating them into semiconductor-based photonic chips, and they might
also have applications in medical devices and treatments, such as spatially selective illumination
of individual neuron cells to locally control neuron firing activities in optogenetics and neuroscience and
near-field high-resolution imaging beyond the optical diffraction limit as well.
\medskip

Theoretically, a big hurdle also exists for studying photon-photon interactions in the strong-coupling regime mainly due to intractable numerical computation for systems with very strong nonlinearity.
The obstacle of nonlinearity in such a system means that any perturbative theories, e.g. using bare electron states
or linear response theory,\,\cite{book} become inadequate for describing both field and electron dynamics in this system.
The presence of an induced polarization, regarded as a source term to the Maxwell equations,\,\cite{r6,apl}
from photo-excited electrons makes it impossible for us to solve the field equations by simply using finite-element analysis\,\cite{fem} or finite-difference-time-domain methods\,\cite{fdtd}.
Although the semiconductor-Bloch equations\,\cite{koch1} and density-matrix equations\,\cite{book,r9},
derived from many-body theory, are able to accurately capture the nonlinear optical response of electrons,
the inclusion of pair scattering effects on both energy relaxation and
optical dephasing precludes an analytical approach for seeking solution of these equations. As a result, there exists only very few theoretical studies\,\cite{koch},
which heavily depend on computer simulation, that focus on simplified one-dimensional strongly-coupled microcavity systems,
in contrast to the three-dimensional structure and self-consistent approach presented in this paper.
\medskip

Physically, not only the high-quality microcavities\,\cite{Scherer} but also the intense surface-plasmon near fields\,\cite{sp1,sp2} could be employed for reaching the strong-coupling goal in semiconductors.
In this paper we solve the self-consistent equations for strongly-coupled electromagnetic-field dynamics and electron quantum kinetics in a quantum dot above the surface of a thick metallic film,
which has not been fully explored so far from either a theoretical or experimental point of view.
This is done based on finding an analytical solution to Green's function\,\cite{aam1,aam2}
for a quantum dot coupled to a semi-infinite metallic material system, which makes it easy to calculate the effect of the induced polarization field as a source term to the Maxwell equations.
In our formalism, the strong light-electron interaction is reflected in the photon-dressed electronic states with a Rabi gap and in the feedback from the induced optical polarization of dressed
electrons to the incident light.
The formalism derived in this paper goes beyond the weak-coupling limit and deals with a much
more realistic structure in the strong-coupling limit for the development of a surface-plasmon polariton laser with a very low threshold pumping.
Our results clearly demonstrate the ability to {\em control probe-field optical gain and absorption switching and photon conversion} by a surface-plasmon field with
temperature-driven frequency detuning in such a nonlinear system led by dressed electron states, very similar to the `gate' control in an optical transistor.
These conclusions should be experimentally observable\,\cite{spie,ol}.
On the other hand, our numerical results also provide an example for demonstrating the so-called quantum plasmonics,\,\cite{natphys} where the nature of surface-plasmon polaritons and the nature of quantum-confined electrons
are hybridized through near-field coupling.
\medskip

In Sec.\,\ref{sec2}, we will introduce our physics model and derive self-consistent equations for determining the coupled scattering dynamics of a surface-plasmon field and the quantum kinetics of electrons in quantum dots.
Section\ \ref{sec3} is devoted to a full discussion of our numerical results, including scattering and optical absorption of surface-plasmon-polariton field by quantum dots, spontaneous emission
and nonlinear optical response of dressed electron states. Some concluding remarks are given in
Sec.\,\ref{sec4}.

\section{Model and Theory}
\label{sec2}

\begin{figure}[p]
\centering
\epsfig{file=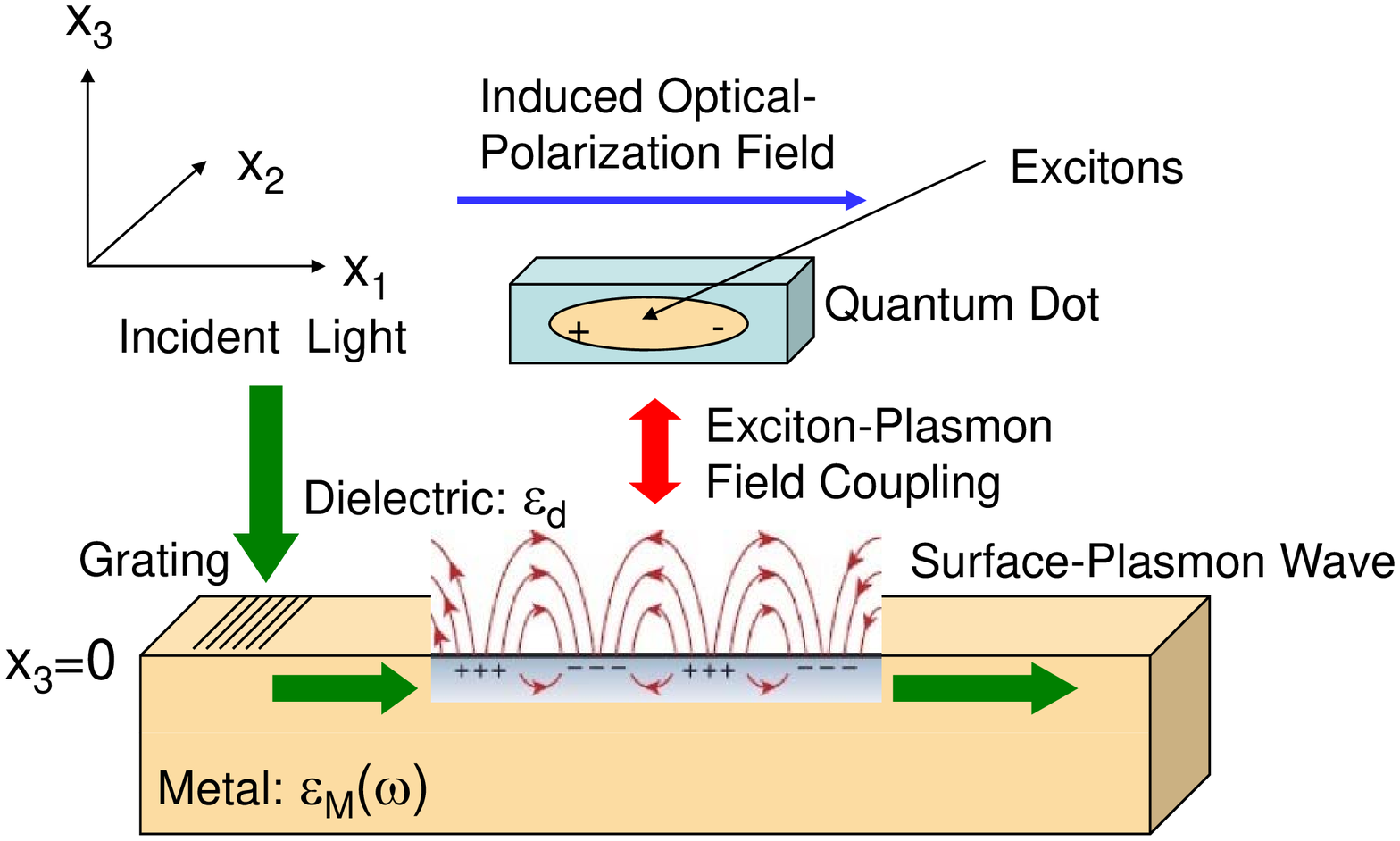,width=0.7\textwidth}
\caption{\label{f1}
(Color online) Schematic illustration for a semi-infinite metal and a quantum dot above its surface at $x_3=0$.
Here, the surface-plasmon polariton (SPP) is locally excited by incident light with the help of a surface grating. The propagating SPP field further excites e-h pairs (plasmas) in the adjacent quantum dot.
As a result, the optical-polarization field of the photo-excited e-h plasma is strongly coupled to the propagating SPP field to
form split plasma-SPP modes with an anticrossing gap. Also, a probe-field is used for studying the photon dressing effect.}
\end{figure}

Our model system, as shown in Fig.\,\ref{f1}, consists of a semi-infinite metallic materialwith a semiconductor quantum dot above its surface.
A surface-plasmon-polariton (SPP) field is locally excited through a surface grating by normally-incident light. This propagating SPP field further excites an
interband electron-hole (e-h) plasma in the quantum dot. The induced optical-polarization field of the photo-excited e-h plasma is strongly coupled to
the SPP field to produce split degenerate e-h plasma and SPP modes with an anticrossing gap.
Part of the brief description for our self-consistent formalism was reported earlier.\,\cite{apl} In order to let readers follow up easily with the details of our model and formalism, we present
here the full derivation of the Maxwell-Bloch numerical approach for an SPP field coupled to a photo-excited e-h plasma in the quantum dot.

\subsection{General Formalism}

The Maxwell's equation for a semi-infinite non-magnetic medium in position-frequency space can be written as\,\cite{aam1}

\begin{equation}
\mbox{\boldmath$\nabla$$\times$\boldmath$\nabla$$\times$\boldmath$E$}({\bf r};\,\omega)-\epsilon_{\rm b}(x_3;\,\omega)\,\frac{\omega^2}{c^2}\,\mbox{\boldmath$E$}({\bf r};\,\omega)
=\frac{\omega^2}{\epsilon_0c^2}\,\mbox{\boldmath${\cal P}$}^{\rm loc}({\bf r};\,\omega)\ ,
\label{e1}
\end{equation}
where $\mbox{\boldmath$E$}({\bf r};\,\omega)$ is the electric component of an electromagnetic field,
$\displaystyle{\mbox{\boldmath$H$}({\bf r};\,\omega)=-\left(\frac{i}{\omega\mu_0}\right)\mbox{\boldmath$\nabla$$\times$\boldmath$E$}({\bf r};\,\omega)}$ is the magnetic component of the electromagnetic field,
${\bf r}=(x_1,x_2,x_3)$ is a three-dimensional position vector, $\omega$ is the angular frequency of the incident light,
$\epsilon_0$, $\mu_0$ and $c$ are the permittivity, permeability and speed of light in vacuum,
$\mbox{\boldmath${\cal P}$}^{\rm loc}({\bf r};\,\omega)$ is an off-surface local polarization field generated
by optical transitions of electrons in a quantum dot, and the position-dependent dielectric function is

\begin{equation}
\epsilon_{\rm b}(x_3;\,\omega)=\left\{\begin{array}{ll}
\epsilon_{\rm d}\ , & \mbox{for $x_3>0$}\\
\epsilon_{\rm M}(\omega)\ , & \mbox{for $x_3<0$}
\end{array}\right.\ .
\label{e2}
\end{equation}
Here, $\epsilon_{\rm d}$ is for the semi-infinite dielectric material in the region $x_3>0$, while $\epsilon_{\rm M}(\omega)$ represents the semi-infinite metallic material in the region $x_3<0$.
For the Maxwell's equation in Eq.\,(\ref{e1}), we introduce the Green's function ${\cal G}_{\mu\nu}({\bf r},{\bf r}^\prime;\,\omega)$ satisfying the following equation

\begin{equation}
\sum\limits_{\mu}\left[\epsilon_{\rm b}(x_3;\,\omega)\,\frac{\omega^2}{c^2}\,\delta_{\lambda\mu}-\frac{\partial^2}{\partial x_{\lambda}\partial x_{\mu}}+\delta_{\lambda\mu}\,\nabla_{\bf r}^2\right]{\cal G}_{\mu\nu}({\bf r},{\bf r}^\prime;\,\omega)=\delta_{\lambda\nu}\,\delta({\bf r}-{\bf r}^\prime)\ ,
\label{e3}
\end{equation}
where $\displaystyle{\nabla_{\bf r}^2=\sum\limits_{\mu}\,\frac{\partial^2}{\partial x^2_{\mu}}}$ is the Laplace operator, $\delta_{\lambda\mu}$ represents the  Kronecker delta, and the indices $\lambda,\,\mu=1,\,2,\,3$ indicate three spatial directions.
Using the Green's function defined in Eq.\,(\ref{e3}),
we can convert the Maxwell's equation in Eq.\,(\ref{e1}) into a three-dimensional integral equation

\begin{equation}
E_{\mu}({\bf r};\,\omega)=E^{(0)}_{\mu}({\bf r};\,\omega)-\frac{\omega^2}{\epsilon_0c^2}\sum\limits_{\nu}\int d^3{\bf r}^\prime\,{\cal G}_{\mu\nu}({\bf r},{\bf r}^\prime;\,\omega)\,
{\cal P}_{\nu}^{\rm loc}({\bf r}^\prime;\,\omega)\ ,
\label{e4}
\end{equation}
where $E^{(0)}_{\mu}({\bf r};\,\omega)$ is a solution of the corresponding homogeneous equation

\begin{equation}
\sum\limits_{\nu}\left[\epsilon_{\rm b}(x_3;\,\omega)\,\frac{\omega^2}{c^2}\,\delta_{\mu\nu}-\frac{\partial^2}{\partial x_{\mu}\partial x_{\nu}}
+\delta_{\mu\nu}\,\nabla^2_{\bf r}\right]E^{(0)}_{\nu}({\bf r};\,\omega)=0\ ,
\label{e5}
\end{equation}
and the source term ${\cal P}_{\nu}^{\rm loc}({\bf r}^\prime;\,\omega)$ generally depends on the electric field in a nonlinear way and can be determined by the Bloch equation.\,\cite{r6,r7}

\subsection{Solving Green's Function}

For a semi-infinite medium, the Green's function can be formally expressed by its Fourier transform

\begin{equation}
{\cal G}_{\mu\nu}({\bf r},{\bf r}^\prime;\,\omega)=\int \frac{d^2{\bf k}_{\|}}{(2\pi)^2}\,e^{i{\bf k}_{\|}\cdot({\bf r}_{\|}-{\bf r}^\prime_{\|})}\,g_{\mu\nu}({\bf k}_{\|},\omega\vert x_3,x_3^\prime)\ ,
\label{e6}
\end{equation}
where we have introduced the notations for the two-dimensional vectors ${\bf r}_{\|}=(x_1,x_2)$ and ${\bf k}_{\|}=(k_1,k_2)$.
Substituting Eq.\,(\ref{e6}) into Eq.\,(\ref{e3}), we obtain

\begin{equation}
\left[\begin{array}{ccc}
\displaystyle{\epsilon_b\,\frac{\omega^2}{c^2}-k_2^2+\frac{d^2}{dx_3^2}} & k_1k_2 & \displaystyle{-ik_1\frac{d}{dx_3}}\\
k_1k_2 & \displaystyle{\epsilon_b\,\frac{\omega^2}{c^2}-k_1^2+\frac{d^2}{dx_3^2}} & \displaystyle{-ik_2\frac{d}{dx_3}}\\
\displaystyle{-ik_1\frac{d}{dx_3}} & \displaystyle{-ik_2\frac{d}{dx_3}} & \displaystyle{\epsilon_b\,\frac{\omega^2}{c^2}-k_{\|}^2}
\end{array}\right]\,\left[\begin{array}{ccc}
g_{11} & g_{12} & g_{13}\\
g_{21} & g_{22} & g_{23}\\
g_{31} & g_{32} & g_{33}
\end{array}\right]=\delta(x_3-x_3^\prime)\,\left[\begin{array}{ccc}
1 & 0 & 0\\
0 & 1 & 0\\
0 & 0 & 1
\end{array}\right]\ .
\label{e7}
\end{equation}
After a rotational transformation\,\cite{aam1} is performed in ${\bf k}_{\|}$-space, i.e.,

\begin{equation}
f_{\mu\nu}(k_{\|},\omega\vert x_3,x_3^\prime)
=\sum\limits_{\mu^\prime,\nu^\prime}\,{\cal S}_{\mu\mu^\prime}({\bf k}_{\|})\,{\cal S}_{\nu\nu^\prime}({\bf k}_{\|})\,g_{\mu^\prime\nu^\prime}({\bf k}_{\|},\omega\vert x_3,x_3^\prime)\ ,
\label{e8}
\end{equation}
where the rotational matrix is selected as

\begin{equation}
{\cal S}({\bf k}_{\|})=\frac{1}{k_{\|}}\,\left[\begin{array}{ccc}
k_1 & k_2 & 0\\
-k_2 & k_1 & 0\\
0 & 0 & k_{\|}
\end{array}\right]\ ,
\label{e9}
\end{equation}
we acquire an equivalent version of Eq.\,(\ref{e7})

\begin{equation}
\left[\begin{array}{ccc}
\displaystyle{\epsilon_b\,\frac{\omega^2}{c^2}+\frac{d^2}{dx_3^2}} & 0 & \displaystyle{-ik_{\|}\frac{d}{dx_3}}\\
0 & \displaystyle{\epsilon_b\,\frac{\omega^2}{c^2}-k_{\|}^2+\frac{d^2}{dx_3^2}} & 0\\
\displaystyle{-ik_{\|}\frac{d}{dx_3}} & 0 & \displaystyle{\epsilon_b\,\frac{\omega^2}{c^2}-k_{\|}^2}
\end{array}\right]\,\left[\begin{array}{ccc}
f_{11} & f_{12} & f_{13}\\
f_{21} & f_{22} & f_{23}\\
f_{31} & f_{32} & f_{33}
\end{array}\right]=\delta(x_3-x_3^\prime)\,\left[\begin{array}{ccc}
1 & 0 & 0\\
0 & 1 & 0\\
0 & 0 & 1
\end{array}\right]\ .
\label{e10}
\end{equation}
To get the solution of Eq.\,(\ref{e10}), we need to employ both the finite-value boundary condition at $x_3^\prime=\pm\infty$ and the continuity boundary condition at the $x_3=0$ interface. This leads to the following
five non-zero $f_{\mu\nu}(k_{\|},\omega\vert x_3,x_3^\prime)$ elements\,\cite{aam1,aam2}

\begin{eqnarray}
&&f_{22}(k_{\|},\omega\vert x_3,x_3^\prime)
\nonumber\\
&=&\left\{\begin{array}{llll}
\displaystyle{-\left(\frac{i}{2p}\right)\frac{2p}{p_{\rm d}+p}\ e^{ip_{\rm d}x_3-ipx_3^\prime}}\ , & x_3>0\ ,\,x_3^\prime<0\\
\\
\displaystyle{-\left(\frac{i}{2p}\right)\left[e^{ip|x_3-x_3^\prime|}-\frac{p_{\rm d}-p}{p_{\rm d}+p}\ e^{-ip(x_3+x_3^\prime)}\right]}\ , & x_3<0\ ,\,x_3^\prime<0\\
\\
\displaystyle{-\left(\frac{i}{2p_{\rm d}}\right)\left[e^{ip_{\rm d}|x_3-x_3^\prime|}+\frac{p_{\rm d}-p}{p_{\rm d}+p}\ e^{ip_{\rm d}(x_3+x_3^\prime)}\right]}\ , & x_3>0\ ,\,x_3^\prime>0\\
\\
\displaystyle{-\left(\frac{i}{2p_{\rm d}}\right)\frac{2p_{\rm d}}{p_{\rm d}+p}\ e^{-ip(x_3-x_3^\prime)}}\ , & x_3<0\ ,\,x_3^\prime>0
\end{array}\right.\ ,\ \
\label{e11}
\end{eqnarray}

\begin{eqnarray}
&&f_{13}(k_{\|},\omega\vert x_3,x_3^\prime)
\nonumber\\
&=&\left\{\begin{array}{llll}
\displaystyle{\frac{ik_{\|}c^2}{2\epsilon_{\rm M}(\omega)\omega^2}\left[\frac{2\epsilon_{\rm M}(\omega)p_{\rm d}}{\epsilon_{\rm M}(\omega)p_{\rm d}+\epsilon_{\rm d}\,p}\right]\,e^{ip_{\rm d}x_3-ipx_3^\prime}}\ , & x_3>0\ ,\,x_3^\prime<0\\
\\
\displaystyle{\frac{ik_{\|}c^2}{2\epsilon_{\rm M}(\omega)\omega^2}\left[e^{ip|x_3-x_3^\prime|}\,{\rm sgn}(x_3-x_3^\prime)+\frac{\epsilon_{\rm M}(\omega)p_{\rm d}-\epsilon_{\rm d}\,p}{\epsilon_{\rm M}(\omega)p_{\rm d}+\epsilon_{\rm d}\,p}\ e^{-ip(x_3+x_3^\prime)}\right]}\ , & x_3<0\ ,\,x_3^\prime<0\\
\\
\displaystyle{\frac{ik_{\|}c^2}{2\epsilon_{\rm d}\,\omega^2}\left[e^{ip_{\rm d}|x_3-x_3^\prime|}\,{\rm sgn}(x_3-x_3^\prime)+\frac{\epsilon_{\rm M}(\omega)p_{\rm d}-\epsilon_{\rm d}\,p}{\epsilon_{\rm M}(\omega)p_{\rm d}+\epsilon_{\rm d}\,p}\ e^{ip_{\rm d}(x_3+x_3^\prime)}\right]}\ , & x_3>0\ ,\,x_3^\prime>0\\
\\
\displaystyle{-\frac{ik_{\|}c^2}{2\epsilon_{\rm d}\,\omega^2}\left[\frac{2\epsilon_{\rm d}\,p}{\epsilon_{\rm M}(\omega)p_{\rm d}+\epsilon_{\rm d}\,p}\right]\,e^{-ipx_3+ip_{\rm d}x_3^\prime}}\ , & x_3<0\ ,\,x_3^\prime>0
\end{array}\right.
\label{e12}
\end{eqnarray}

\begin{eqnarray}
&&f_{33}(k_{\|},\omega\vert x_3,x_3^\prime)
\nonumber\\
&=&\left\{\begin{array}{llll}
\displaystyle{-\frac{ik^2_{\|}c^2}{\omega^2}\left[\frac{1}{\epsilon_{\rm M}(\omega)p_{\rm d}+\epsilon_{\rm d}\,p}\right]\,e^{ip_{\rm d}x_3-ipx_3^\prime}}\ , & x_3>0\ ,\,x_3^\prime<0\\
\\
\displaystyle{\frac{c^2}{\epsilon_{\rm M}(\omega)\omega^2}\,\delta(x_3-x_3^\prime)-\frac{ik^2_{\|}c^2}{2p\,\epsilon_{\rm M}(\omega)\omega^2}\left[e^{ip|x_3-x_3^\prime|}-\frac{\epsilon_{\rm M}(\omega)p_{\rm d}-\epsilon_{\rm d}\,p}{\epsilon_{\rm M}(\omega)p_{\rm d}+\epsilon_{\rm d}\,p}\ e^{-ip(x_3+x_3^\prime)}\right]}\ , & x_3<0\ ,\,x_3^\prime<0\\
\\
\displaystyle{\frac{c^2}{\epsilon_{\rm d}\,\omega^2}\,\delta(x_3-x_3^\prime)-\frac{ik_{\|}^2c^2}{2p_{\rm d}\epsilon_{\rm d}\,\omega^2}\left[e^{ip_{\rm d}|x_3-x_3^\prime|}+\frac{\epsilon_{\rm M}(\omega)p_{\rm d}-\epsilon_{\rm d}\,p}{\epsilon_{\rm M}(\omega)p_{\rm d}+\epsilon_{\rm d}\,p}\ e^{ip_{\rm d}(x_3+x_3^\prime)}\right]}\ , & x_3>0\ ,\,x_3^\prime>0\\
\\
\displaystyle{-\frac{ik^2_{\|}c^2}{\omega^2}\left[\frac{1}{\epsilon_{\rm M}(\omega)p_{\rm d}+\epsilon_{\rm d}\,p}\right]\,e^{-ipx_3+ip_{\rm d}x_3^\prime}}\ , & x_3<0\ ,\,x_3^\prime>0
\end{array}\right.
\label{e13}
\end{eqnarray}

\begin{eqnarray}
&&f_{11}(k_{\|},\omega\vert x_3,x_3^\prime)
\nonumber\\
&=&\left\{\begin{array}{llll}
\displaystyle{-\frac{ip_{\rm d}p\,c^2}{\omega^2}\left[\frac{1}{\epsilon_{\rm M}(\omega)p_{\rm d}+\epsilon_{\rm d}\,p}\right]\,e^{ip_{\rm d}x_3-ipx_3^\prime}}\ , & x_3>0\ ,\,x_3^\prime<0\\
\\
\displaystyle{-\frac{ip\,c^2}{2\epsilon_{\rm M}(\omega)\omega^2}\left[e^{ip|x_3-x_3^\prime|}+\frac{\epsilon_{\rm M}(\omega)p_{\rm d}-\epsilon_{\rm d}\,p}{\epsilon_{\rm M}(\omega)p_{\rm d}+\epsilon_{\rm d}\,p}\ e^{-ip(x_3+x_3^\prime)}\right]}\ , & x_3<0\ ,\,x_3^\prime<0\\
\\
\displaystyle{-\frac{ip_{\rm d}c^2}{2\epsilon_{\rm d}\,\omega^2}\left[e^{ip_{\rm d}|x_3-x_3^\prime|}-\frac{\epsilon_{\rm M}(\omega)p_{\rm d}-\epsilon_{\rm d}\,p}{\epsilon_{\rm M}(\omega)p_{\rm d}+\epsilon_{\rm d}\,p}\ e^{ip_{\rm d}(x_3+x_3^\prime)}\right]}\ , & x_3>0\ ,\,x_3^\prime>0\\
\\
\displaystyle{-\frac{ip_{\rm d}c^2}{2\epsilon_{\rm d}\,\omega^2}\left[\frac{2\epsilon_{\rm d}p}{\epsilon_{\rm M}(\omega)p_{\rm d}+\epsilon_{\rm d}\,p}\right]\,e^{-ipx_3+ip_{\rm d}x_3^\prime}}\ , & x_3<0\ ,\,x_3^\prime>0
\end{array}\right.
\label{e14}
\end{eqnarray}

\begin{eqnarray}
&&f_{31}(k_{\|},\omega\vert x_3,x_3^\prime)
\nonumber\\
&=&\left\{\begin{array}{llll}
\displaystyle{\frac{ik_{\|}c^2}{\omega^2}\left[\frac{p}{\epsilon_{\rm M}(\omega)p_{\rm d}+\epsilon_{\rm d}\,p}\right]\,e^{ip_{\rm d}x_3-ipx_3^\prime}}\ , & x_3>0\ ,\,x_3^\prime<0\\
\\
\displaystyle{\frac{ik_{\|}c^2}{2\epsilon_{\rm M}(\omega)\omega^2}\left[e^{ip|x_3-x_3^\prime|}\,{\rm sgn}(x_3-x_3^\prime)-\frac{\epsilon_{\rm M}(\omega)p_{\rm d}-\epsilon_{\rm d}\,p}{\epsilon_{\rm M}(\omega)p_{\rm d}+\epsilon_{\rm d}\,p}\ e^{-ip(x_3+x_3^\prime)}\right]}\ , & x_3<0\ ,\,x_3^\prime<0\\
\\
\displaystyle{\frac{ik_{\|}c^2}{2\epsilon_{\rm d}\,\omega^2}\left[e^{ip_{\rm d}|x_3-x_3^\prime|}\,{\rm sgn}(x_3-x_3^\prime)-\frac{\epsilon_{\rm M}(\omega)p_{\rm d}-\epsilon_{\rm d}\,p}{\epsilon_{\rm M}(\omega)p_{\rm d}+\epsilon_{\rm d}\,p}\ e^{ip_{\rm d}(x_3+x_3^\prime)}\right]}\ , & x_3>0\ ,\,x_3^\prime>0\\
\\
\displaystyle{-\frac{ik_{\|}c^2}{\omega^2}\left[\frac{p_{\rm d}}{\epsilon_{\rm M}(\omega)p_{\rm d}+\epsilon_{\rm d\,}p}\right]\,e^{-ipx_3+ip_{\rm d}x_3^\prime}}\ , & x_3<0\ ,\,x_3^\prime>0
\end{array}\right.
\label{e15}
\end{eqnarray}
where ${\rm sgn}(x)$ is the sign function,

\begin{equation}
p_{\rm d}(k_{\|},\omega)=\sqrt{\epsilon_{\rm d}\,\frac{\omega^2}{c^2}-k_{\|}^2}\ ,
\label{e16}
\end{equation}

\begin{equation}
p(k_{\|},\omega)=\sqrt{\epsilon_{\rm M}(\omega)\,\frac{\omega^2}{c^2}-k_{\|}^2}\ ,
\label{e17}
\end{equation}
${\rm Im}[p_{\rm d}(k_{\|},\omega)]\geq 0$ and ${\rm Im}[p(k_{\|},\omega)]\geq 0$.
In addition, from these non-zero $f_{\mu\nu}(k_{\|},\omega\vert x_3,x_3^\prime)$ functions, we obtain

\begin{equation}
g_{\mu\nu}({\bf k}_{\|},\omega\vert x_3,x_3^\prime)
=\sum\limits_{\mu^\prime,\nu^\prime}\,f_{\mu^\prime\nu^\prime}(k_{\|},\omega\vert x_3,x_3^\prime)\,{\cal S}_{\mu^\prime\mu}({\bf k}_{\|})\,{\cal S}_{\nu^\prime\nu}({\bf k}_{\|})\ ,
\label{e18}
\end{equation}
which can be substituted into Eq.\,(\ref{e6}) to calculate the Green's function ${\cal G}_{\mu\nu}({\bf r},{\bf r}^\prime;\,\omega)$ in position space.

\subsection{Local Polarization Field}

In order to find the explicit field dependence in $\mbox{\boldmath${\cal P}$}^{\rm loc}({\bf r};\,\omega)$, we now turn to the study of electron dynamics in a quantum dot.
Here, the optical-polarization field $\mbox{\boldmath${\cal P}$}^{\rm loc}({\bf r};\,\omega)$ plays a unique role on bridging the classical Maxwell's equations
for electromagnetic fields to the quantum-mechanical Schr\"odinger equation for electrons.
The electron dynamics in photo-excited quantum dots can be described quantitatively by the so-called semiconductor Bloch equations\,\cite{r3,r4,r5}. These generalize the well-known
optical Bloch equations in two aspects including the incorporation of electron scattering with impurities, phonons and other electrons as well as
many-body effects on dephasing in the photo-induced optical coherence.
\medskip

\begin{figure}[p]
\centering
\epsfig{file=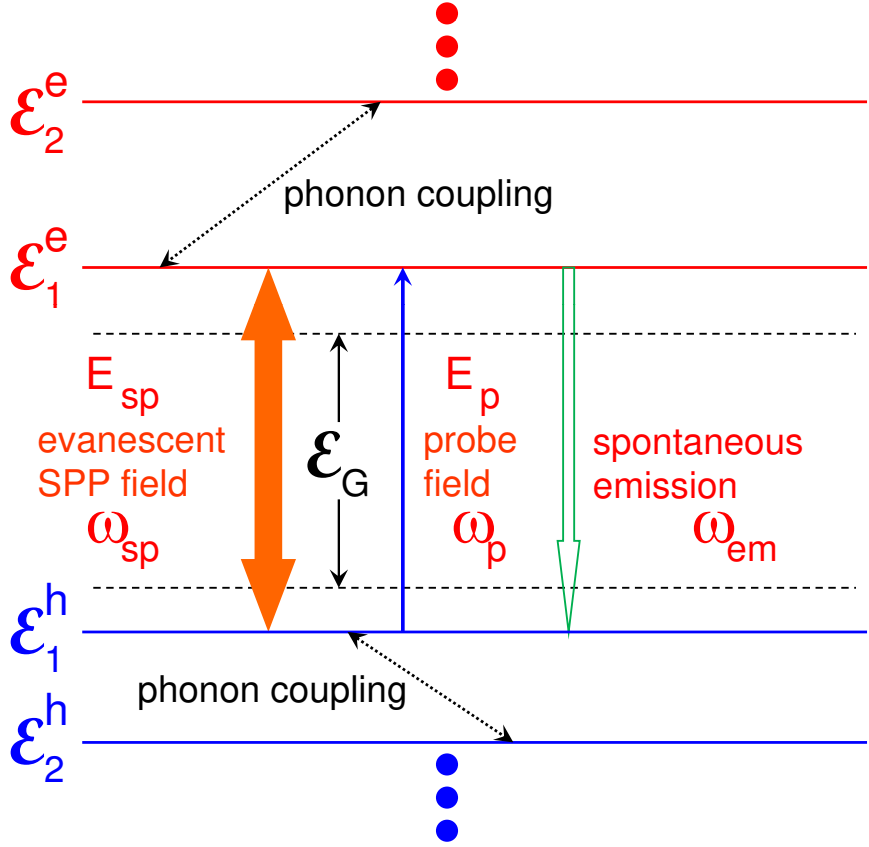,width=0.7\textwidth}
\caption{\label{f2}
(Color online) Schematic illustration for a system incorporating the generation of quantum-dot excitons
by a SPP field with frequency $\omega_{\rm sp}$ and probed by a plane-wave field $E_p$ with frequency $\omega_p$. Here, ${\cal E}_{\rm G}$ is the
energy bandgap of the host semiconductor, $\varepsilon^{\rm e}_\ell$
and $\varepsilon^{\rm h}_j$ stand for the energy levels of electrons
and holes, respectively, with $\ell,\,j=1,\,2,\,\cdots$. In addition, $\omega_{em}$ represents the frequency of spontaneous emission from photo-excited excitons,
and the ground states of electrons and holes are coupled to their first excited states by lattice phonons at finite temperatures.}
\end{figure}

The physical system considered in this paper is illustrated in Fig.\,\ref{f2}, where we assume two levels for electrons and holes, respectively, in a quantum dot.
These two energy levels of both electrons and holes are efficiently coupled by phonon scattering at high temperatures. Additionally, the lowest electron and hole energy levels are optically coupled
to each other by an incident SPP field to form the dressed states of excitons. The SPP-controlled optical properties of quantum-dot excitons can either probed by a plane-wave field or seen from the
spontaneous emission of excitons.
\medskip

For photo-excited spin-degenerated electrons in the conduction band, the semiconductor Bloch equations with $\ell=1,\,2,\,\cdots$ are given by

\begin{equation}
\frac{dn^{\rm e}_\ell}{dt}=\frac{2}{\hbar}\,\sum\limits_j\,{\rm Im}\left[\left(Y_\ell^j\right)^\ast\left({\cal M}^{\rm eh}_{\ell,j}-Y_\ell^j\,V^{\rm
eh}_{\ell,j;j,\ell}\right)\right]
+\left.\frac{\partial n^{\rm e}_\ell}{\partial t}\right|_{\rm rel}-\delta_{\ell,1}\,{\cal R}_{\rm sp}\,n_1^{\rm e}\,n_1^{\rm h}\ ,
\label{e19}
\end{equation}
where ${\cal R}_{\rm sp}$ is the spontaneous emission rate and $n^{\rm e}_\ell$ represents the electron level population. In Eq.\,(\ref{e19}), the term marked `rel' is the non-radiative energy relaxation
for $n^{\rm e}_\ell$, and the $Y_\ell^j$, ${\cal M}^{\rm eh}_{\ell,j}$, and $V^{\rm eh}_{\ell,j;j,\ell}$ terms are given later in the text.
\medskip

Similarly, for spin-degenerate holes in the valence band, the semiconductor Bloch equations with $j=1,\,2,\,\cdots$ are found to be

\begin{equation}
\frac{dn^{\rm h}_j}{dt}=\frac{2}{\hbar}\,\sum\limits_\ell\,{\rm Im}\left[\left(Y_\ell^j\right)^\ast\left({\cal M}^{\rm eh}_{\ell,j}-Y_\ell^j\,V^{\rm
eh}_{\ell,j;j,\ell}\right)\right]+\left.\frac{\partial n^{\rm h}_j}{\partial t}\right|_{\rm rel}-\delta_{j,1}\,{\cal R}_{\rm sp}\,n_1^{\rm e}\,n_1^{\rm h}\ ,
\label{e20}
\end{equation}
where $n^{\rm h}_j$ stands for the hole energy level population. Again, the non-radiative energy relaxation for $n^{\rm h}_j$ is incorporated in Eq.\,(\ref{e20}). Moreover, we know from Eqs.\,(\ref{e19}) and (\ref{e20}) that

\begin{equation}
N_{\rm e}(t)=2\,\sum\limits_\ell\,n^{\rm e}_\ell(t)=2\,\sum\limits_j\,n^{\rm h}_j(t)=N_{\rm h}(t)\ ,
\label{e21}
\end{equation}
where $N_{\rm e}(t)$ and $N_{\rm h}(t)$ are the total number of photo-excited electrons and holes, respectively, in the quantum dot at time $t$.
\medskip

Finally, for spin-averaged e-h plasmas, the induced interband optical coherence, which is introduced in Eqs.\,(\ref{e19}) and (\ref{e20}), with $j=1,\,2,\,\cdots$ and $\ell=1,\,2,\,\cdots$ satisfies the following equations,

\[
i\hbar\,\frac{d}{dt}Y_\ell^j=\left[\overline{\varepsilon}^{\rm
e}_\ell(\omega)+\overline{\varepsilon}^{\rm
h}_j(\omega)-\hbar(\omega+i\gamma_0)\right]Y_\ell^j+\left(1-n^{\rm
e}_\ell-n^{\rm h}_j\right)\left({\cal M}^{\rm eh}_{\ell,j}-Y_\ell^j\,V^{\rm
eh}_{\ell,j;j,\ell}\right)
\]
\[
+Y_\ell^j\left[\sum\limits_{j_1}\,n^{\rm h}_{j_1}\left(V^{\rm hh}_{j,j_1;j_1,j}-V^{\rm hh}_{j,j_1;j,j_1}\right)-\sum\limits_{\ell_1}\,n^{\rm e}_{\ell_1}\,V^{\rm eh}_{\ell_1,j;j,\ell_1}\right]
\]
\begin{equation}
+Y_\ell^j\left[\sum\limits_{\ell_1}\,n^{\rm e}_{\ell_1}\left(V^{\rm ee}_{\ell,\ell_1;\ell_1,\ell}-V^{\rm ee}_{\ell,\ell_1;\ell,\ell_1}\right)-\sum\limits_{j_1}\,n^{\rm h}_{j_1}\,V^{\rm eh}_{\ell,j_1;j_1,\ell}\right]\ ,
\label{e22}
\end{equation}
where $\hbar\gamma_0=\hbar\gamma_{\rm eh}+\hbar\gamma_{ext}$ is the total energy-level broadening due to both
the finite carrier lifetime and the loss of an external evanescent field, $\omega$ is the frequency of the external field, and
$\overline{\varepsilon}^{\rm e}_\ell(\omega)$ and
$\overline{\varepsilon}^{\rm h}_j(\omega)$ are the kinetic energies
of dressed single electrons and holes, respectively (see Appendix\ A with
$\alpha=1$). In Eq.\,(\ref{e22}), the diagonal dephasing ($\gamma_0$) of
$Y^j_\ell$, the renormalization of interband Rabi coupling ($Y_\ell^j\,V^{\rm
eh}_{\ell,j;j,\ell}$), the
renormalization of electron and hole energies (third and fourth terms on the right-hand side), as well as the
exciton binding energy, are all taken into consideration. Since the
e-h plasmas are independent of spin index in this case, they can be
excited by both left-circularly and right-circularly polarized light.
The off-diagonal dephasing of $Y^j_\ell$ has been neglected due to low carrier density.
\medskip

The steady-state solution to Eq.\,(\ref{e22}), i.e. under the condition of $dY^j_\ell/dt=0$, is found to be

\begin{equation}
Y^j_\ell(t\vert\omega)=\left[\frac{1-n^{\rm
e}_\ell(t)-n_j^{\rm
h}(t)}{\hbar(\omega+i\gamma_0)-\hbar\overline{\Omega}^{\rm
eh}_{\ell,j}(\omega\vert t)}\right]{\cal M}^{\rm eh}_{\ell,j}(t)\ ,
\label{e23}
\end{equation}
where the photon and Coulomb renormalized interband energy-level separation $\hbar\overline{\Omega}^{\rm eh}_{\ell,j}(\omega\vert t)$ is given by

\[
\hbar\overline{\Omega}^{\rm eh}_{\ell,j}(\omega\vert t)=\overline{\varepsilon}^{\rm e}_\ell(\omega\vert t)+\overline{\varepsilon}^{\rm h}_j(\omega\vert t)-V^{eh}_{\ell,j;j,\ell}+\sum\limits_{\ell_1}\,n^{\rm e}_{\ell_1}(t)\left(V^{\rm ee}_{\ell,\ell_1;\ell_1,\ell}-V^{\rm ee}_{\ell,\ell_1;\ell,\ell_1}\right)
\]
\begin{equation}
+\sum\limits_{j_1}\,n^{\rm h}_{j_1}(t)\left(V^{\rm hh}_{j,j_1;j_1,j}-V^{\rm hh}_{j,j_1;j,j_1}\right)-\sum\limits_{\ell_1\neq \ell}\,n^{\rm e}_{\ell_1}(t)\,V^{\rm eh}_{\ell_1,j;j,\ell_1}-\sum\limits_{j_1\neq j}\,n^{\rm h}_{j_1}(t)\,V^{\rm eh}_{\ell,j_1;j_1,\ell}\ .
\label{e24}
\end{equation}
The steady-state solution in Eq.\,(\ref{e23}) can be substituted into Eqs.\,(\ref{e19}) and (\ref{e20}) above.
\medskip

The Coulomb interaction matrix elements introduced in Eqs.(\ref{e19}), (\ref{e20}) and (\ref{e22}) are defined as

\[
V^{\rm ee}_{\ell_1,\ell_2;\ell_3,\ell_4}=\int d^3{\bf r}\int d^3{\bf
r}^\prime\,\left[\psi^{\rm e}_{\ell_1}({\bf
r})\right]^\ast\left[\psi^{\rm e}_{\ell_2}({\bf
r}^\prime)\right]^\ast\frac{e^2}{4\pi\epsilon_0\epsilon_b|{\bf
r}-{\bf r}^\prime|}\,\psi^{\rm e}_{\ell_3}({\bf
r}^\prime)\,\psi^{\rm e}_{\ell_4}({\bf r})
\]
\begin{equation}
=\frac{e^2}{8\pi^2\epsilon_0\epsilon_b}\int d^2{\bf q}_\|\,{\cal
F}^{\rm e}_{\ell_1,\ell_4}({\bf q}_\|)\,{\cal F}^{\rm
e}_{\ell_2,\ell_3}(-{\bf q}_\|)\left(\frac{1}{q_\|+q_s}\right)
=\left(V^{\rm ee}_{\ell_1,\ell_2;\ell_3,\ell_4}\right)^\ast\ ,
\label{e25}
\end{equation}

\[
V^{\rm hh}_{j_1,j_2;j_3,j_4}=\int d^3{\bf r}\int d^3{\bf
r}^\prime\,\left[\psi^{\rm h}_{j_1}({\bf
r})\right]^\ast\left[\psi^{\rm h}_{j_2}({\bf
r}^\prime)\right]^\ast\frac{e^2}{4\pi\epsilon_0\epsilon_b|{\bf
r}-{\bf r}^\prime|}\,\psi^{\rm h}_{j_3}({\bf r}^\prime)\,\psi^{\rm
h}_{j_4}({\bf r})
\]
\begin{equation}
=\frac{e^2}{8\pi^2\epsilon_0\epsilon_b}\int d^2{\bf q}_\|\,{\cal
F}^{\rm h}_{j_1,j_4}({\bf q}_\|)\,{\cal F}^{\rm h}_{j_2,j_3}(-{\bf
q}_\|)\left(\frac{1}{q+q_s}\right)=\left(V^{\rm
hh}_{j_1,j_2;j_3,j_4}\right)^\ast\ , \label{e26}
\end{equation}

\[
V^{\rm eh}_{\ell,j;j^\prime,\ell^\prime}=\int d^3{\bf r}\int d^3{\bf r}^\prime\,\left[\psi^{\rm e}_{\ell}({\bf r})\right]^\ast\left[\psi^{\rm h}_{j}({\bf r}^\prime)\right]^\ast\frac{e^2}{4\pi\epsilon_0\epsilon_b|{\bf r}-{\bf r}^\prime|}\,\psi^{\rm h}_{j^\prime}({\bf r}^\prime)\,\psi^{\rm e}_{\ell^\prime}({\bf r})
\]
\begin{equation}
=\frac{e^2}{8\pi^2\epsilon_0\epsilon_b}\int d^2{\bf q}_\|\,{\cal
F}^{\rm e}_{\ell,\ell^\prime}({\bf q}_\|)\,{\cal F}^{\rm
h}_{j,j^\prime}(-{\bf
q}_\|)\left(\frac{1}{q+q_s}\right)=\left(V^{\rm
eh}_{\ell,j;j^\prime,\ell^\prime}\right)^\ast\ , \label{e27}
\end{equation}
where the static screening length $1/q_s$ at temperatures ($k_BT\gg E_F$) is determined from

\begin{equation}
q_s(t)=\frac{e^2}{4\epsilon_0\epsilon_b{\cal S}\,k_{\rm B}T}\left[N_{\rm e}(t)+N_{\rm h}(t)\right]\ ,
\label{e28}
\end{equation}
${\cal S}$ is the cross-sectional area of a quantum dot, $T$ is the lattice temperature, $\psi^{\rm e}_\ell({\bf r})$ and $\psi^{\rm h}_j({\bf r})$ are the envelope wave-functions of electrons and holes in a quantum dot (see Appendix\ A), $\epsilon_b$ is the average dielectric constant of the host semiconductor. The two dimensionless form factors (see Appendix\ A) introduced in Eqs.\,(\ref{e25})-(\ref{e27}) for electrons and holes due to quantum confinement by a quantum dot are defined by

\begin{equation}
{\cal F}_{\ell,\ell^\prime}^{\rm e}({\bf q_\|})=e^{-q_\|{\cal L}_0}\int d^2{\bf
r}_\|\left[\psi_\ell^{\rm e}({\bf r}_\|)\right]^\ast{\rm e}^{i{\bf
q}_\|\cdot{\bf r}_\|}\,\psi_{\ell^\prime}^{\rm e}({\bf r}_\|)=\left[{\cal
F}_{\ell,\ell^\prime}^{\rm e}(-{\bf q}_\|)\right]^\ast\ , \label{e29}
\end{equation}

\begin{equation}
{\cal F}_{j,j^\prime}^{\rm h}({\bf q}_\|)=e^{-q_\|{\cal L}_0}\int d^2{\bf
r}_\|\left[\psi_j^{\rm h}({\bf r}_\|)\right]^\ast{\rm e}^{i{\bf
q}_\|\cdot{\bf r}_\|}\,\psi_{j^\prime}^{\rm e}({\bf r}_\|)=\left[{\cal
F}_{j,j^\prime}^{\rm h}(-{\bf q}_\|)\right]^\ast\ , \label{e30}
\end{equation}
where ${\cal L}_0$ is the thickness of a disk-like quantum dot.
In addition, the matrix elements employed in Eqs.\,(\ref{e19}), (\ref{e20}) and (\ref{e22}) for the Rabi coupling between photo-excited carriers and an evanescent external field $\displaystyle{\mbox{\boldmath$E$}({\bf r};\,t)=\theta(t)\,\mbox{\boldmath$E$}({\bf r};\,\omega)\,e^{-i\omega t}}$ are given by

\begin{equation}
{\cal M}^{\rm eh}_{\ell,j}(t)=-\delta_{\ell,1}\,\delta_{j,1}\,\theta(t)\,\left[\mbox{\boldmath$E$}^{\rm eh}_{\ell,j}(\omega)\cdot{\bf d}_{\rm c,v}\right]\ ,
\label{e31}
\end{equation}
where $\theta(x)$ is a unit step function, the static interband dipole moment ${\bf d}_{\rm c,v}$ (see Appendix\ A) is

\begin{equation}
{\bf d}_{\rm c,v}=\int d^3{\bf r}\left[u_{\rm c}({\bf
r})\right]^\ast\,{\bf r}\,u_{\rm v}({\bf r})={\bf d}_{\rm c,v}^\ast\ ,
\label{e32}
\end{equation}
$u_{\rm c}({\bf r})$ and $u_{\rm v}({\bf r})$ are the Bloch functions associated with conduction and valence bands at the $\Gamma$-point in the first Brillouin zone of the host semiconductor, and the effective electric field coupled to the quantum dot is

\begin{equation}
\mbox{\boldmath$E$}^{\rm eh}_{\ell,j}(\omega)=\int d^3{\bf r}\left[\psi^{\rm e}_\ell({\bf r})\right]^\ast\mbox{\boldmath$E$}({\bf r};\,\omega)\left[\psi^{\rm h}_j({\bf r})\right]^\ast\ .
\label{e33}
\end{equation}
\medskip

The Boltzmann-type scattering term\,\cite{force} for non-radiative electron energy relaxation in Eq.\,(\ref{e19}) is

\begin{equation}
\left.\frac{\partial n^{\rm e}_\ell}{\partial t}\right|_{\rm rel}={\cal W}_\ell^{({\rm in})}(1-n_\ell^{\rm e})-{\cal W}_\ell^{({\rm out})}\,n_\ell^{\rm e}\ ,
\label{e34}
\end{equation}
where the microscopic scattering-in and scattering-out rates are calculated as

\[
{\cal W}_\ell^{({\rm in})}=\frac{2\pi}{\hbar}\,\sum\limits_{\ell^\prime}\,^\prime\,\left|V^{\rm ep}_{\ell,\ell^\prime}\right|^2\,n^{\rm e}_{\ell^\prime}\left\{{\cal N}_{\rm ph}(\Omega_0)\left[\frac{\hbar\Gamma_{\rm ph}/\pi}{(\overline{\varepsilon}^{\rm e}_\ell-\overline{\varepsilon}^{\rm e}_{\ell^\prime}-\hbar\Omega_0)^2+\hbar^2\Gamma^2_{\rm ph}}\right]\right.
\]
\[
\left.+\left[{\cal N}_{\rm ph}(\Omega_0)+1\right]\left[\frac{\hbar\Gamma_{\rm ph}/\pi}{(\overline{\varepsilon}^{\rm e}_\ell-\overline{\varepsilon}^{\rm e}_{\ell^\prime}+\hbar\Omega_0)^2+\hbar^2\Gamma^2_{\rm ph}}\right]\right\}
\]
\begin{equation}
+\frac{2\pi}{\hbar}\,\sum\limits_{\ell^\prime}\,^\prime\,\sum\limits_{j,j^\prime}\,^\prime\,\left|V^{\rm eh}_{\ell,j;j^\prime,\ell^\prime}\right|^2\,(1-n_j^{\rm h})\,n^{\rm h}_{j^\prime}\,n^{\rm e}_{\ell^\prime}\left[\frac{\hbar\gamma_{\rm eh}/\pi}{(\overline{\varepsilon}^{\rm e}_{\ell}+\overline{\varepsilon}^{\rm h}_{j}-\overline{\varepsilon}^{\rm e}_{\ell^\prime}-\overline{\varepsilon}^{\rm h}_{j^\prime})^2+\hbar^2\gamma_{\rm eh}^2}\right]\ ,
\label{e35}
\end{equation}

\[
{\cal W}_\ell^{({\rm out})}=\frac{2\pi}{\hbar}\,\sum\limits_{\ell^\prime}\,^\prime\,\left|V^{\rm ep}_{\ell,\ell^\prime}\right|^2\,(1-n^{\rm e}_{\ell^\prime})\left\{{\cal N}_{\rm ph}(\Omega_0)\left[\frac{\hbar\Gamma_{\rm ph}/\pi}{(\overline{\varepsilon}^{\rm e}_{\ell^\prime}-\overline{\varepsilon}^{\rm e}_{\ell}-\hbar\Omega_0)^2+\hbar^2\Gamma^2_{\rm ph}}\right]\right.
\]
\[
\left.+\left[{\cal N}_{\rm ph}(\Omega_0)+1\right]\left[\frac{\hbar\Gamma_{\rm ph}/\pi}{(\overline{\varepsilon}^{\rm e}_{\ell^\prime}-\overline{\varepsilon}^{\rm e}_{\ell}+\hbar\Omega_0)^2+\hbar^2\Gamma^2_{\rm ph}}\right]\right\}
\]
\begin{equation}
+\frac{2\pi}{\hbar}\,\sum\limits_{\ell^\prime}\,^\prime\,\sum\limits_{j,j^\prime}\,^\prime\,\left|V^{\rm eh}_{\ell^\prime,j;j^\prime,\ell}\right|^2\,(1-n^{\rm e}_{\ell^\prime})\,(1-n^{\rm h}_{j})\,n_{j^\prime}^{\rm h}\left[\frac{\hbar\gamma_{\rm eh}/\pi}{(\overline{\varepsilon}^{\rm e}_{\ell^\prime}+\overline{\varepsilon}^{\rm h}_{j}-\overline{\varepsilon}^{\rm e}_{\ell}-\overline{\varepsilon}^{\rm h}_{j^\prime})^2+\hbar^2\gamma_{\rm eh}^2}\right]\ .
\label{e36}
\end{equation}
Here,the primed summations in Eqs.\,(\ref{e35}) and (\ref{e36}) exclude the terms satisfying either $j=j^\prime$ or $\ell^\prime=\ell$, ${\cal N}_{\rm ph}(\Omega_0)=[\exp(\hbar\Omega_0/k_{\rm B}T)-1]^{-1}$ is the Bose function for the thermal-equilibrium phonons, and $\Omega_0$ and $\Gamma_{\rm ph}$ are the frequency and lifetime of longitudinal-optical phonons in the host semiconductor. Similarly, the Boltzmann-type scattering term for hole non-radiative energy relaxation in Eq.\,(\ref{e20}) is

\begin{equation}
\left.\frac{\partial n^{\rm h}_j}{\partial t}\right|_{\rm rel}=\overline{{\cal W}}_j^{({\rm in})}(1-n_j^{\rm h})-\overline{{\cal W}}_j^{({\rm out})}\,n_j^{\rm h}\ ,
\label{e37}
\end{equation}
where the scattering-in and scattering-out rates are

\[
\overline{{\cal W}}_j^{({\rm in})}=\frac{2\pi}{\hbar}\,\sum\limits_{j^\prime}\,^\prime\,\left|V^{\rm hp}_{j,j^\prime}\right|^2\,n^{\rm h}_{j^\prime}\left\{{\cal N}_{\rm ph}(\Omega_0)\left[\frac{\hbar\Gamma_{\rm ph}/\pi}{(\overline{\varepsilon}^{\rm h}_j-\overline{\varepsilon}^{\rm h}_{j^\prime}-\hbar\Omega_0)^2+\hbar^2\Gamma^2_{\rm ph}}\right]\right.
\]
\[
\left.+\left[{\cal N}_{\rm ph}(\Omega_0)+1\right]\left[\frac{\hbar\Gamma_{\rm ph}/\pi}{(\overline{\varepsilon}^{\rm h}_j-\overline{\varepsilon}^{\rm h}_{j^\prime}+\hbar\Omega_0)^2+\hbar^2\Gamma^2_{\rm ph}}\right]\right\}
\]
\begin{equation}
+\frac{2\pi}{\hbar}\,\sum\limits_{\ell,\ell^\prime}\,^\prime\,\sum\limits_{j^\prime}\,^\prime\,\left|V^{\rm eh}_{\ell,j;j^\prime,\ell^\prime}\right|^2\,(1-n_\ell^{\rm e})\,n^{\rm h}_{j^\prime}\,n^{\rm e}_{\ell^\prime}\left[\frac{\hbar\gamma_{\rm eh}/\pi}{(\overline{\varepsilon}^{\rm e}_{\ell}+\overline{\varepsilon}^{\rm h}_{j}-\overline{\varepsilon}^{\rm e}_{\ell^\prime}-\overline{\varepsilon}^{\rm h}_{j^\prime})^2+\hbar^2\gamma_{\rm eh}^2}\right]\ ,
\label{e38}
\end{equation}

\[
\overline{{\cal W}}_j^{({\rm out})}=\frac{2\pi}{\hbar}\,\sum\limits_{j^\prime}\,^\prime\,\left|V^{\rm hp}_{j,j^\prime}\right|^2\,(1-n^{\rm h}_{j^\prime})\left\{{\cal N}_{\rm ph}(\Omega_0)\left[\frac{\hbar\Gamma_{\rm ph}/\pi}{(\overline{\varepsilon}^{\rm h}_{j^\prime}-\overline{\varepsilon}^{\rm h}_{j}-\hbar\Omega_0)^2+\hbar^2\Gamma^2_{\rm ph}}\right]\right.
\]
\[
\left.+\left[{\cal N}_{\rm ph}(\Omega_0)+1\right]\left[\frac{\hbar\Gamma_{\rm ph}/\pi}{(\overline{\varepsilon}^{\rm h}_{j^\prime}-\overline{\varepsilon}^{\rm h}_{j}+\hbar\Omega_0)^2+\hbar^2\Gamma^2_{\rm ph}}\right]\right\}
\]
\begin{equation}
+\frac{2\pi}{\hbar}\,\sum\limits_{\ell,\ell^\prime}\,^\prime\,\sum\limits_{j^\prime}\,^\prime\,\left|V^{\rm eh}_{\ell,j^\prime;j,\ell^\prime}\right|^2\,(1-n^{\rm e}_{\ell})\,(1-n^{\rm h}_{j^\prime})\,n_{\ell^\prime}^{\rm e}\left[\frac{\hbar\gamma_{\rm eh}/\pi}{(\overline{\varepsilon}^{\rm e}_{\ell}+\overline{\varepsilon}^{\rm h}_{j^\prime}-\overline{\varepsilon}^{\rm e}_{\ell^\prime}-\overline{\varepsilon}^{\rm h}_{j})^2+\hbar^2\gamma_{\rm eh}^2}\right]\ ,
\label{e39}
\end{equation}
and again the primed summations in Eqs.\,(\ref{e38}) and (\ref{e39}) exclude the terms satisfying either $j^\prime=j$ or $\ell=\ell^\prime$. The coupling between the longitudinal-optical phonons and electrons or holes in Eqs.\,(\ref{e35}), (\ref{e36}), (\ref{e38}) and (\ref{e39}) are calculated as

\begin{equation}
\left|V^{\rm ep}_{\ell,\ell^\prime}\right|^2=\frac{e^2\hbar\Omega_0}{8\pi^2\epsilon_0}\left(\frac{1}{\epsilon_\infty}-\frac{1}{\epsilon_s}\right)\int d^2{\bf q}_\|\,\left|{\cal F}^{\rm e}_{\ell,\ell^\prime}({\bf q}_\|)\right|^2\left(\frac{1}{q_\|+q_s}\right)\ ,
\label{e40}
\end{equation}

\begin{equation}
\left|V^{\rm hp}_{\ell,\ell^\prime}\right|^2=\frac{e^2\hbar\Omega_0}{8\pi^2\epsilon_0}\left(\frac{1}{\epsilon_\infty}-\frac{1}{\epsilon_s}\right)\int d^2{\bf q}_\|\,\left|{\cal F}^{\rm h}_{j,j^\prime}({\bf q}_\|)\right|^2\left(\frac{1}{q_\|+q_s}\right)\ ,
\label{e41}
\end{equation}
where $\epsilon_\infty$ and $\epsilon_s$ are the high-frequency and static dielectric constants of the host polar semiconductor.
\medskip

By generalizing the Kubo-Martin-Schwinger relation,\,\cite{r9} the time-dependent spontaneous emission rate, ${\cal R}_{\rm sp}(t)$, introduced in Eqs.\,(\ref{e19}) and (\ref{e20}), can be expressed as

\[
{\cal R}_{\rm sp}(t)=\frac{\left|{\bf d}^\prime_{\rm
c,v}(t)\right|^2}{\epsilon_0\sqrt{\epsilon_b}}\left|\int d^3{\bf
r}\,\psi^{\rm e}_1({\bf r})\,\psi^{\rm h}_1({\bf r})\right|^2
\int\limits_0^\infty
d\omega^\prime\,\theta\left[\hbar\omega^\prime-{\cal E}_{\rm
c}(t)-\overline{\varepsilon}^{\rm
e}_1(\omega\vert t)-\overline{\varepsilon}^{\rm
h}_1(\omega\vert t)\right]
\]
\begin{equation}
\times\hbar\omega^\prime\,\rho_0(\omega^\prime)\left\{\frac{\hbar\gamma_{\rm eh}}{[\hbar\omega^\prime-{\cal
E}_{\rm c}(t)-\overline{\varepsilon}^{\rm
e}_1(\omega\vert t)-\overline{\varepsilon}^{\rm
h}_1(\omega\vert t)]^2+\hbar^2\gamma^2_{\rm eh}}\right\}\ ,
\label{e42}
\end{equation}
where

\begin{equation}
\left|{\bf d}^\prime_{\rm c,v}(t)\right|^2=\frac{e^2\hbar^2}{2m_0\,{\cal E}_{\rm G}(T)}\left[1+\frac{{\cal E}_{\rm c}(t)}{{\cal E}_{\rm G}(T)}\right]\left(\frac{m_0}{m_{\rm e}^\ast}-1\right)\ ,
\label{e43}
\end{equation}
${\cal E}_{\rm G}(T)={\cal E}_{\rm G}(0)-5.41\times 10^{-4}\,T^2/(T+204)$ (in units of eV) is the energy bandgap of the host semiconductor, $\rho_0(\omega)=\omega^2/c^3\pi^2\hbar$ is the density-of-states of spontaneously-emitted photons in vacuum, $m_0$ is the free electron mass, $m_{\rm e}^\ast$ is the effective mass of electrons, and the Coulomb renormalization of the energy bandgap ${\cal E}_{\rm c}(t)$ is calculated as

\[
{\cal E}_{\rm c}(t)=\sum\limits_{\ell_1}\,n^{\rm e}_{\ell_1}(t)\left(V^{\rm ee}_{1,\ell_1;\ell_1,1}-V^{\rm ee}_{1,\ell_1;1,\ell_1}\right)+\sum\limits_{j_1}\,n^{\rm h}_{j_1}(t)\left(V^{\rm hh}_{1,j_1;j_1,1}-V^{\rm hh}_{1,j_1;1,j_1}\right)
\]
\begin{equation}
-\sum\limits_{\ell_1}\,n^{\rm e}_{\ell_1}(t)\,V^{\rm eh}_{\ell_1,1;1,\ell_1}-\sum\limits_{j_1}\,n^{\rm h}_{j_1}(t)\,V^{\rm eh}_{1,j_1;j_1,1}
-\left[1-n^{\rm e}_1(t)-n^{\rm h}_1(t)\right]\,V^{\rm eh}_{1,1;1,1}\ .
\label{e44}
\end{equation}
In Eq.\,(\ref{e44}), the first two terms are associated with the Hartree-Fock energies of electrons and holes, while the rest of the terms are related to the exciton binding energy.
\medskip

Finally, the photo-induced interband optical polarization $\mbox{\boldmath${\cal P}$}^{\rm loc}({\bf r};\,\omega)$,
which is related to the induced interband optical coherence, by dressed electrons in the quantum dot is given by\,\cite{r7}

\[
\mbox{\boldmath${\cal P}$}^{\rm loc}({\bf r};\,\omega)=2\left|\xi({\bf
r})\right|^2\,{\bf d}_{\rm c,v}\left\{\int d^3{\bf r}^\prime\,\psi^{\rm e}_1({\bf
r}^\prime)\,\psi^{\rm h}_1({\bf r}^\prime)\right\}
\]
\begin{equation}
\times\,\frac{1}{\hbar}\,\lim\limits_{t\to\infty}\left[\frac{1-n^{\rm e}_1(t)-n_1^{\rm
h}(t)}{\omega+i\gamma_0-\overline{\Omega}^{\rm
eh}_{1,1}(\omega\vert t)}\right]{\cal M}^{\rm eh}_{1,1}(t)\ , \label{e45}
\end{equation}
where ${\bf d}_{\rm c,v}=d_{\rm c,v}\,\hat{\bf e}_{\rm d}$ represents the interband dipole moment [see Eq.\,(\ref{e32})],
$\hat{\bf e}_{\rm d}$ is the unit vector of the dipole moment, and $|\xi({\bf r})|^2$ comes from the confinement of a quantum dot.

\subsection{Self-Consistent Field Equation}

Since the wavelength of the incident light is much larger than the size of a quantum dot, we can treat the quantum dot, which is excited resonantly by the incident light, as a
point dipole at ${\bf r}={\bf r}_0=(0,0,z_0)$, i.e. we can assume $\mbox{\boldmath${\cal P}$}^{\rm loc}({\bf r}^\prime;\,\omega)=\mbox{\boldmath${\cal P}$}^{\rm loc}(\omega)\,\delta({\bf r}^\prime-{\bf r}_0)$ in Eq.\,(\ref{e4})
to neglect its geometry effect. This greatly simplifies the calculation and gives rise to

\begin{equation}
E_{\mu}({\bf r};\,\omega)=E^{(0)}_{\mu}({\bf r};\,\omega)-\frac{\omega^2}{\epsilon_0c^2}\sum\limits_{\nu}\,{\cal G}_{\mu\nu}({\bf r},{\bf r}_0;\,\omega)\,
{\cal P}_{\nu}^{\rm loc}(\omega)\ ,
\label{e46}
\end{equation}
where

\[
\mbox{\boldmath${\cal P}$}^{\rm loc}(\omega)=2{\bf d}_{\rm
c,v}\left\{\int d^3{\bf r}^\prime\,\psi^{\rm e}_1({\bf
r}^\prime)\,\psi^{\rm h}_1({\bf r}^\prime)\right\}
\]
\begin{equation}
\times\frac{1}{\hbar}\,\lim\limits_{t\to\infty}\left\{\frac{1-n^{\rm e}_1(t)-n_1^{\rm
h}(t)}{\omega+i\gamma_0-\overline{\Omega}^{\rm
eh}_{1,1}(\omega\vert t)}\right\}{\cal M}^{\rm eh}_{1,1}(t)\ ,
\label{e47}
\end{equation}

\begin{equation}
{\cal M}^{\rm eh}_{1,1}(t)=-\theta(t)\left[\mbox{\boldmath$E$}({\bf r}_0;\,\omega)\cdot{\bf d}_{\rm c,v}\right]
\left\{\int d^3{\bf r}\,\psi^{\rm e}_1({\bf r})\,\psi^{\rm h}_1({\bf r})\right\}^\ast\ .
\label{e48}
\end{equation}
Substituting Eqs.\,(\ref{e47}) and (\ref{e48}) into Eq.\,(\ref{e46}), we get the following nonlinear equations for the electromagnetic field

\[
E_{\mu}({\bf r};\,\omega)=E^{(0)}_{\mu}({\bf r};\,\omega)+\frac{2\omega^2}{\epsilon_0c^2}
\left[\mbox{\boldmath$E$}({\bf r}_0;\,\omega)\cdot{\bf d}_{\rm c,v}\right]\,d_{\rm c,v}\,
\left|\int d^3{\bf r}^\prime\,\psi^{\rm e}_1({\bf
r}^\prime)\,\psi^{\rm h}_1({\bf r}^\prime)\right|^2
\]
\begin{equation}
\times\frac{1}{\hbar}\,\lim\limits_{t\to\infty}\left\{\frac{1-n^{\rm e}_1(t)-n_1^{\rm
h}(t)}{\omega+i\gamma_0-\overline{\Omega}^{\rm
eh}_{1,1}(\omega\vert t)}\right\}
\sum\limits_{\nu}\,{\cal G}_{\mu\nu}({\bf r},{\bf r}_0;\,\omega)\,\hat{e}_{\rm d}^{\nu}\ ,
\label{e49}
\end{equation}
where the quantum-dot level populations $n^{\rm e}_\ell(t)$ and $n_j^{\rm h}(t)$ depend nonlinearly on $\mbox{\boldmath$E$}({\bf r}_0;\,\omega)$ in the strong-coupling regime.
\medskip

If the electromagnetic field is not very strong, we can neglect the pumping effect. In this linear-response regime,
we can write down the electron and hole populations in a thermal-equilibrium state [without solving Eqs.\,(\ref{e19}) and (\ref{e20})]

\begin{equation}
n^{\rm e}_\ell(t)\approx f_0(\overline{\varepsilon}^{\rm e}_{\ell})\equiv\frac{1}{\exp[(\overline{\varepsilon}^{\rm e}_{\ell}-\mu_{\rm e})/k_BT]+1}\ ,
\label{e50}
\end{equation}

\begin{equation}
n^{\rm h}_j(t)\approx f_0(\overline{\varepsilon}^{\rm h}_j)\equiv\frac{1}{\exp[(\overline{\varepsilon}^{\rm h}_j-\mu_{\rm h})/k_BT]+1}\ ,
\label{e51}
\end{equation}
where $f_0(x)$ is the Fermi function, $\mu_{\rm e}$ and $\mu_{\rm h}$ are the chemical potentials of electrons and holes, respectively,
determined by Eq.\,(\ref{e21}). As a result of Eqs.\,(\ref{e50}) and (\ref{e51}), we get from Eq.\,(\ref{e49}) the {\em linearized} self-consistent field equation at ${\bf r}={\bf r}_0$

\begin{equation}
\sum\limits_{\nu}\,{\cal A}_{\mu\nu}({\bf r}_0;\,\omega)\,
E_{\nu}({\bf r}_0;\,\omega)=E^{(0)}_{\mu}({\bf r}_0;\,\omega)
\label{e52}
\end{equation}
with
\[
{\cal A}_{\mu\nu}({\bf r}_0;\,\omega)=\delta_{\mu\nu}-\frac{2\omega^2}{\epsilon_0c^2\hbar}
\left[\frac{1-f_0(\overline{\varepsilon}^{\rm e}_1)-f_0(\overline{\varepsilon}^{\rm h}_1)}{\omega+i\gamma_0-\overline{\Omega}^{\rm
eh}_{1,1}(\omega)}\right]\,\left|\int d^3{\bf r}^\prime\,\psi^{\rm e}_1({\bf
r}^\prime)\,\psi^{\rm h}_1({\bf r}^\prime)\right|^2
\]
\begin{equation}
\times d^2_{\rm c,v}\left[\hat{e}_{\rm d}^{\nu}\sum\limits_{\nu_1}\,{\cal G}_{\mu\nu_1}({\bf r}_0,{\bf r}_0;\,\omega)\,\hat{e}_{\rm d}^{\nu_1}\right]\ ,
\end{equation}
\label{e53}
where, according to Eq.\,(\ref{e6}), we have

\begin{equation}
{\cal G}_{\mu\nu}({\bf r}_0,{\bf r}_0;\,\omega)=\int \frac{d^2{\bf k}_{\|}}{(2\pi)^2}\,g_{\mu\nu}({\bf k}_{\|},\omega\vert z_0,z_0)\ .
\label{e54}
\end{equation}
The solution $\mbox{\boldmath$E$}({\bf r}_0;\,\omega)$ of the linear-matrix equation in Eq.\,(\ref{e52}) can be substituted into Eq.\,(\ref{e49}) to find the spatial distribution of the electromagnetic field $\mbox{\boldmath$E$}({\bf r};\,\omega)$ at all positions other than ${\bf r}={\bf r}_0$, i.e.,

\[
E_{\mu}({\bf r};\,\omega)=E^{(0)}_{\mu}({\bf r};\,\omega)+\frac{2\omega^2}{\epsilon_0c^2\hbar}
\left[\sum\limits_{\nu,\nu^\prime}\,\hat{e}_{\rm d}^{\nu}\,{\cal A}^{-1}_{\nu\nu^\prime}({\bf r}_0;\,\omega)\,E^{(0)}_{\nu^\prime}({\bf r}_0;\,\omega)\right]
\]
\begin{equation}
\times\left|\int d^3{\bf r}^\prime\,\psi^{\rm e}_1({\bf
r}^\prime)\,\psi^{\rm h}_1({\bf r}^\prime)\right|^2\,d^2_{\rm c,v}\left[\frac{1-f_0(\overline{\varepsilon}^{\rm e}_1)-f_0(\overline{\varepsilon}^{\rm h}_1)}{\omega+i\gamma_0-\overline{\Omega}^{\rm
eh}_{1,1}(\omega)}\right]\,
\sum\limits_{\nu_1}\,{\cal G}_{\mu\nu_1}({\bf r},{\bf r}_0;\,\omega)\,\hat{e}_{\rm d}^{\nu_1}\ .
\label{e55}
\end{equation}
\medskip

In order to find the coupled e-h plasma and plasmon dispersion relation $\omega=\Omega_{\rm ex-pl}({\bf k}_{\|})$, we perform Fourier transforms to both $\mbox{\boldmath$E$}({\bf r};\,\omega)$ and $\mbox{\boldmath$E$}^{(0)}({\bf r};\,\omega)$ in Eq.\,(\ref{e46}) with respect to ${\bf r}_{\|}$. This leads to

\begin{equation}
E_{\mu}({\bf k}_{\|},\omega\vert x_3)=E^{(0)}_{\mu}({\bf k}_{\|},\omega\vert x_3)-\frac{\omega^2}{\epsilon_0c^2}\sum\limits_{\nu}\,g_{\mu\nu}({\bf k}_{\|},\omega\vert x_3,z_0)\,
{\cal P}_{\nu}^{\rm loc}(\omega)\ .
\label{e56}
\end{equation}
After setting $x_3=z_0$ in Eq.\,(\ref{e56}), we get

\begin{eqnarray}
&&\sum\limits_{\nu}\left\{\delta_{\mu\nu}-\frac{2\omega^2}{\epsilon_0c^2\hbar}\,
\left[\frac{1-f_0(\overline{\varepsilon}^{\rm e}_1)-f_0(\overline{\varepsilon}^{\rm h}_1)}{\omega+i\gamma_0-\overline{\Omega}^{\rm
eh}_{1,1}(\omega)}\right]\,\left|\int d^3{\bf r}^\prime\,\psi^{\rm e}_1({\bf
r}^\prime)\,\psi^{\rm h}_1({\bf r}^\prime)\right|^2\,d^2_{\rm c,v}\right.
\nonumber\\
&\times&\left.\left[\hat{e}_{\rm d}^{\nu}\sum\limits_{\nu_1}\,g_{\mu\nu_1}({\bf k}_{\|},\omega\vert z_0,z_0)\,\hat{e}_{\rm d}^{\nu_1}\right]\right\}E_{\nu}({\bf k}_{\|},\omega\vert z_0)=E^{(0)}_{\mu}({\bf k}_{\|},\omega\vert z_0)\ .
\label{e57}
\end{eqnarray}
Here, the zero determinant of the coefficient matrix in Eq.\,(\ref{e57}) determines the coupled e-h plasma and plasmon dispersion relation $\omega=\Omega_{\rm ex-pl}({\bf k}_{\|})$.
We emphasize that the assumption of thermal-equilibrium states for electrons and holes is just for obtaining analytical expressions. Therefore, some qualitative conclusions can be drawn
for guidance from these analytical solutions. Our numerical results, however, are based on the non-thermal-equilibrium steady states calculated after solving self-consistently the coupled Maxwell-Bloch equations.
\medskip

By assuming an incident SPP field within the $x_1x_2$-plane, we can write

\begin{equation}
\mbox{\boldmath$E$}^{(0)}({\bf r};\,\omega_{\rm sp})=E_{\rm sp}\,e^{i{\bf k}_0(\omega_{\rm sp})\cdot{\bf D}_0}\,\frac{c}{\omega_{\rm sp}}
\left[i\hat{\bf k}_0\beta_3(k_0,\omega_{\rm sp})-\hat{\bf x}_3k_0(\omega_{\rm sp})\right]\,e^{i{\bf k}_0(\omega_{\rm sp})\cdot{\bf x}_\|}\,e^{-\beta_3(k_0,\,\omega_{\rm sp})x_3}\ ,
\label{e58}
\end{equation}
where ${\bf x}_\|=\{x_1,\,x_2\}$, $\hat{\bf k}_0$ and $\hat{\bf x}_3$ are the unit vectors in the ${\bf k}_0=k_0(\omega_{\rm sp})\{\cos\theta_0,\,\sin\theta_0\}$
and $x_3$ directions, $E_{\rm sp}$ is the field amplitude, $\omega_{\rm sp}$ is the field frequency, $\theta_0$ is the angle of the incident SPP field with respect to the $x_1$ direction, ${\bf D}_0=\{-x_g,\,-y_g\}$ is the position vector
of the surface grating, and the two wave numbers are

\begin{equation}
k_0(\omega_{\rm sp})=\frac{\omega_{\rm sp}}{c}\sqrt{\frac{\epsilon_{\rm d}\,\epsilon_{\rm M}(\omega_{\rm sp})}{\epsilon_{\rm d}+\epsilon_{\rm M}(\omega_{\rm sp})}}\ ,
\label{e59}
\end{equation}

\begin{equation}
\beta_3(k_0,\omega_{\rm sp})=\sqrt{k^2_0(\omega_{\rm sp})-\frac{\omega_{\rm sp}^2}{c^2}}\ ,
\label{e60}
\end{equation}
with ${\rm Re}[k_0(\omega_{\rm sp})]\geq 0$ and ${\rm Re}[\beta_3(k_0,\omega_{\rm sp})]\geq 0$.
Here, the in-plane wave number $k_0$ is produced by the surface-grating diffraction of the $p$-polarized normally-incident light, which in turn determines the resonant frequency $\omega$ of the SPP mode.
Equation (\ref{e59}) stands for the full dispersion relation of the SPP field, including both radiative and non-radiative parts.
From Eq.\,(\ref{e58}), it is easy to find its Fourier transformed expression

\[
\mbox{\boldmath$E$}^{(0)}({\bf k}_{\|},\omega_{\rm sp}\vert z_0)=\delta({\bf k}_{\|}-{\bf k}_0)\,E_{\rm sp}\,e^{i{\bf k}_0(\omega_{\rm sp})\cdot{\bf D}_0}\,\frac{(2\pi)^2c}{\omega_{\rm sp}}
\]
\begin{equation}
\times\left[i\hat{\bf k}_0\beta_3(k_0,\omega_{\rm sp})-\hat{\bf x}_3k_0(\omega_{\rm sp})\right]\,e^{-\beta_3(k_0,\,\omega_{\rm sp})z_0}\ .
\label{e61}
\end{equation}

\subsection{Quantum-Dot Absorption}

On the basis of the above electromagnetic field $\mbox{\boldmath$E$}({\bf r}_0;\,\omega)$ at the quantum dot, we are able to compute the time-resolved nonlinear interband absorption coefficient
of {\em dressed} electrons in a quantum dot for the SPP field.\,\cite{r11} In this case, we find

\begin{equation}
\beta_0(\omega;\,t)=\frac{\omega\sqrt{\epsilon_b}}{n_{\rm spp}(\omega;\,t)\,c}\left[\frac{1}{\exp(\hbar\omega/k_{\rm
B}T)-1}+1\right]\,{\rm Im}\left[\alpha_{\rm spp}(\omega;\,t)\right]\ ,
\label{add1}
\end{equation}
where $\alpha_{\rm spp}(\omega;\,t)$ is the complex Lorentz function given by

\[
{\rm Im}[\alpha_{\rm spp}(\omega;\,t)]=\theta(t)\left(\frac{2}{\epsilon_0\epsilon_b{\cal
V}|\mbox{\boldmath$E$}^{(0)}({\bf r}_0;\,\omega)|^2}\right)\left|\mbox{\boldmath$E$}({\bf r}_0;\,\omega)\cdot{\bf d}_{\rm
c,v}\right|^2\,\left|\int d^3{\bf r}\,\psi^{\rm
e}_1({\bf r})\,\psi^{\rm h}_1({\bf r})\right|^2
\]
\begin{equation}
\times\left[1-n^{\rm e}_1(t)-n_1^{\rm h}(t)\right]\left\{\frac{[A^2(\omega;\,t)-B^2(t)]^2+4\hbar^2\gamma_0^2A^2(\omega;\,t)}
{[A^2(\omega;\,t)+B^2(t)]^2+4\hbar^2\gamma_0^2A^2(\omega;\,t)}\right\}\,\left[\frac{\hbar\gamma_0}{\Delta^2(\omega;\,t)+\hbar^2\gamma_0^2}\right]\ ,
\label{add2}
\end{equation}

\[
{\rm Re}[\alpha_{\rm spp}(\omega;\,t)]=-\theta(t)\left(\frac{2}{\epsilon_0\epsilon_b{\cal
V}|\mbox{\boldmath$E$}^{(0)}({\bf r}_0;\,\omega)|^2}\right)\left|\mbox{\boldmath$E$}({\bf r}_0;\,\omega)\cdot{\bf d}_{\rm
c,v}\right|^2\,\left|\int d^3{\bf r}\,\psi^{\rm
e}_1({\bf r})\,\psi^{\rm h}_1({\bf r})\right|^2
\]
\begin{equation}
\times\left[1-n^{\rm e}_1(t)-n_1^{\rm h}(t)\right]\left\{\frac{A^4(\omega;\,t)-B^4(t)]^2}
{[A^2(\omega;\,t)+B^2(t)]^2}\right\}\,\left[\frac{\Delta(\omega;\,t)}{\Delta^2(\omega;\,t)+\hbar^2\gamma_0^2}\right]\ ,
\label{add3}
\end{equation}
and the scaled refractive index function $n_{\rm spp}(\omega;\,t)$ can be calculated by

\[
n_{\rm spp}(\omega;\,t)=
\frac{1}{\sqrt{2}}\left(1+{\rm Re}\left[\alpha_{\rm spp}(\omega;\,t)\right]\right.
\]
\begin{equation}
\left.+\sqrt{\left\{1+{\rm Re}\left[\alpha_{\rm spp}(\omega;\,t)\right]\right\}^2+\left\{{\rm Im}\left[\alpha_{\rm spp}
(\omega;\,t)\right]\right\}^2}\,\right)^{1/2}\ .
\label{add4}
\end{equation}
In Eqs.\,(\ref{add2}) and (\ref{add3}), the dressed-state effects on both the level population and dipole moment have been included.
In addition, we have introduced the following notations in Eqs.\,(\ref{add2})
and (\ref{add3})

\begin{equation}
\Delta(\omega;\,t)=\sqrt{[{\cal E}_{\rm
G}(T)+\varepsilon_1^{\rm e}+\varepsilon_1^{\rm
h}-\hbar\omega]^2+4|{\cal M}^{\rm
eh}_{1,1}(t)|^2}\ ,
\label{add5}
\end{equation}

\begin{equation}
A^2(\omega;\,t)=\left[\hbar\omega-{\cal E}_{\rm G}(T)-\varepsilon_1^{\rm e}-\varepsilon_1^{\rm h}+\Delta(\omega;\,t)\right]^2\ ,\ \ \ \ B^2(t)=4|{\cal M}^{\rm eh}_{1,1}(t)|^2\ .
\label{add6}
\end{equation}

\subsection{Probing Quantum-Dot Dressed States}

We are also able to compute the time-resolved linear interband absorption coefficient of electrons, dressed by the SPP field, for a weak probe field (not the strong SPP field) on the basis of the above calculated electromagnetic field $\mbox{\boldmath$E$}({\bf r}_0;\,\omega)$ at the quantum dot.\,\cite{r11}
Assuming a spatially-uniform probe field $\displaystyle{\mbox{\boldmath$E$}_p(t)=\theta(t-\tau)\,\mbox{\boldmath$E$}_p\,e^{-i\omega_pt}}$ with $\tau$ being the delay time, the probe-field absorption coefficient
$\beta_{\rm abs}(\omega_p;\,t)$ of the lowest dressed state is given by Eq.\,(\ref{add1}) with the replacements of $\omega$, $n_{\rm spp}$, and $\alpha_{\rm spp}$ by
$\omega_p$, $n_{\rm pf}$, and $\alpha_{\rm pf}$, respectively,
where

\[
\alpha_{\rm pf}(\omega_p;\,t)=-\theta(t-\tau)\left(\frac{2}{\epsilon_0\epsilon_b{\cal
V}|\mbox{\boldmath$E$}_p|^2\hbar}\right)\left|\mbox{\boldmath$E$}_p\cdot{\bf d}_{\rm
c,v}\right|^2\,\left|\int d^3{\bf r}\,\psi^{\rm
e}_1({\bf r})\,\psi^{\rm h}_1({\bf
r})\right|^2\,\left[1-n_1^{\rm e}(t)-n_1^{\rm h}(t)\right]
\]
\begin{equation}
\times\left\{\frac{A^2(\omega;\,t)-B^2(t)}{[A^2(\omega;\,t)+B^2(t)]^2}\right\}
\left\{\frac{A^2(\omega;\,t)}{\omega_p+i\gamma_{\rm eh}-\overline{\Omega}^{\rm
eh}_{1,1}(\omega_-\vert t)}-\frac{B^2(t)}{\omega_p+i\gamma_{\rm eh}-\overline{\Omega}^{\rm
eh}_{1,1}(\omega_+\vert t)}\right\}\ ,
\label{e63}
\end{equation}

\begin{eqnarray}
&&n_{\rm pf}(\omega_p;\,t)=
\nonumber\\
&&\frac{1}{\sqrt{2}}\left(1+{\rm Re}\left[\alpha_{\rm
pf}(\omega_p;\,t)\right]+\sqrt{\left\{1+{\rm Re}\left[\alpha_{\rm
pf}(\omega_p;\,t)\right]\right\}^2+\left\{{\rm Im}\left[\alpha_{\rm
pf}(\omega_p;\,t)\right]\right\}^2}\,\right)^{1/2}\ . \label{e64}
\end{eqnarray}
Here, using Eq.\,(\ref{e24}) we have

\[
\hbar\overline{\Omega}^{\rm eh}_{1,1}(\omega_\pm\vert t)=
\hbar\omega_{\pm}(t)-\left[1-n^{\rm e}_1(t)-n^{\rm
h}_1(t)\right]V^{\rm eh}_{1,1;1,1}+\sum\limits_{\ell_1}\,n^{\rm
e}_{\ell_1}(t)\left(V^{\rm ee}_{1,\ell_1;\ell_1,1}-V^{\rm
ee}_{1,\ell_1;1,\ell_1}\right)
\]
\begin{equation}
+\sum\limits_{j_1}\,n^{\rm h}_{j_1}(t)\left(V^{\rm
hh}_{1,j_1;j_1,1}-V^{\rm
hh}_{1,j_1;1,j_1}\right)-\sum\limits_{\ell_1}\,n^{\rm
e}_{\ell_1}(t)\,V^{\rm
eh}_{\ell_1,1;1,\ell_1}-\sum\limits_{j_1}\,n^{\rm
h}_{j_1}(t)\,V^{\rm eh}_{1,j_1;j_1,1}\ ,
\label{e65}
\end{equation}
and

\begin{equation}
\hbar\omega_{\pm}(t)=\hbar\omega\pm\Delta(\omega;\,t)\ .
\label{e66}
\end{equation}
\medskip

Moreover, the time-resolved photoluminescence spectrum ${\cal P}_{\rm em}(\omega^\prime;\,t)$ of dressed electrons in the quantum dot is proportional to

\[
{\cal P}_{\rm em}(\omega^\prime;\,t)\propto\frac{|{\bf d}^\prime_{\rm
c,v}|^2}{\epsilon_0\sqrt{\epsilon_b}{\cal L}_0}\,n_1^{\rm e}(t)\,n_1^{\rm h}(t)\,\hbar\gamma_{\rm eh}\,\left\{\frac{1}{[A^2(\omega;\,t)+B^2(t)]^2}\right\}\,
\left|\int d^3{\bf r}\,\psi^{\rm
e}_1({\bf r})\,\psi^{\rm h}_1({\bf
r})\right|^2\,\hbar\omega^\prime\,\rho_0(\omega^\prime)
\]
\[
\times\left\{\frac{A^2(\omega;\,t)\,B^2(t)}{[\hbar\omega^\prime-{\cal
E}_{\rm c}(t)-\hbar\omega_-(t)]^2+\hbar^2\gamma^2_{\rm eh}}
+\frac{A^2(\omega;\,t)\,B^2(t)}{[\hbar\omega^\prime-{\cal E}_{\rm c}(t)-\hbar\omega_+(t)]^2+\hbar^2\gamma^2_{\rm eh}}\right.
\]
\begin{equation}
+\left.\frac{A^4(\omega;\,t)+B^4(t)}{[\hbar\omega^\prime-{\cal
E}_{\rm c}(t)-\hbar\omega]^2+\hbar^2\gamma^2_{\rm eh}}\right\}\ .
\label{e69}
\end{equation}

\section{Numerical Results and Discussions}
\label{sec3}

\subsection{Results for the dynamics of an SPP field}

In the first part of our numerical calculations, we have taken: ${\cal L}_0=100$\,\AA, $L_y=100$\,\AA, $m^\ast_{\rm e}=0.067\,m_0$, $m^\ast_{\rm h}=0.62\,m_0$, $\theta_0=45^{\rm o}$, $x_g=y_g=610$\,\AA, $\epsilon_b=12$,
$\epsilon_s=11$, $\epsilon_{\infty}=13$, $\hbar\Omega_0=36$\,meV and $\hbar\gamma_{\rm eh}=\hbar\Gamma_{\rm ph}=\hbar\gamma_0$.
The silver plasma frequency is $13.8\times10^{15}$\,Hz and the silver plasma dephasing is $0.1075\times10^{15}$\,Hz. The energy gap $E_{\rm G}$ of the quantum-dot host material is $1.927$\,eV at $T=300$\,K.
Other parameters, including $T$, $E_{\rm sp}$, $L_x$, $\hbar\gamma_0$, $z_0$ and $\epsilon_{\rm d}$, will be directly indicated in the figures.
\medskip

\begin{figure}[p]
\centering
\epsfig{file=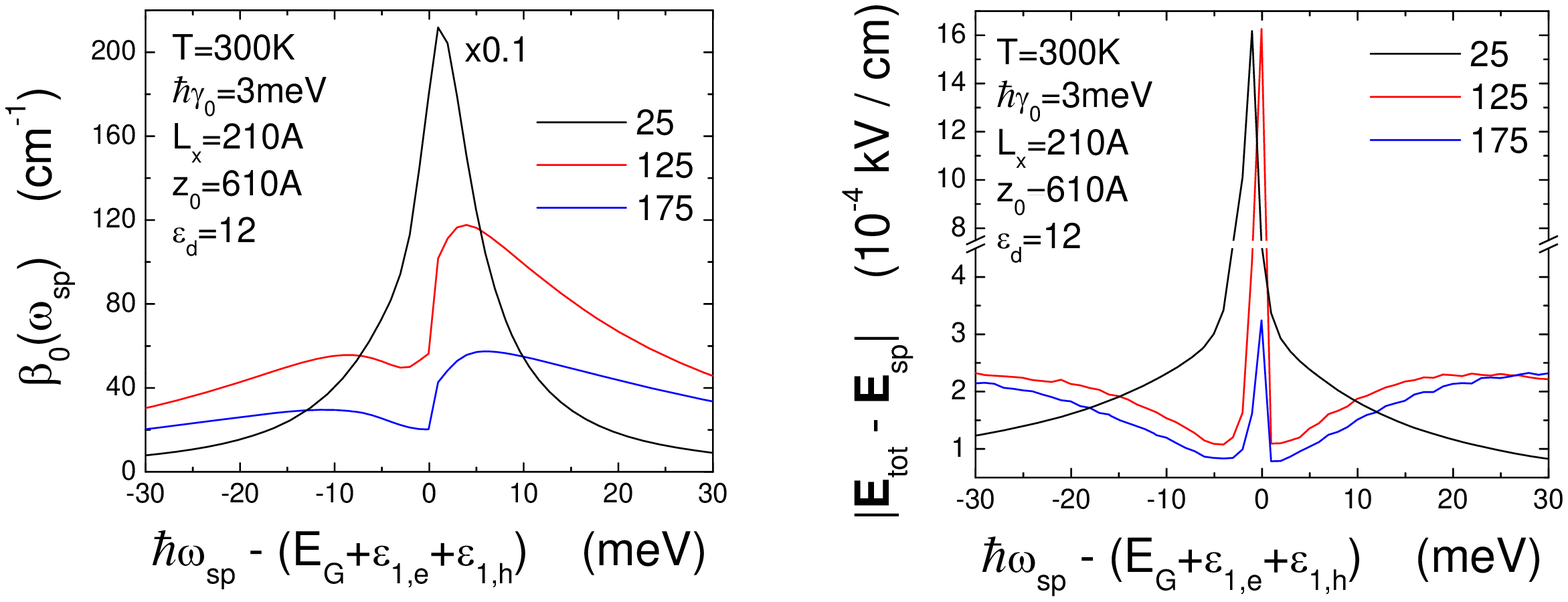,width=0.85\textwidth}
\epsfig{file=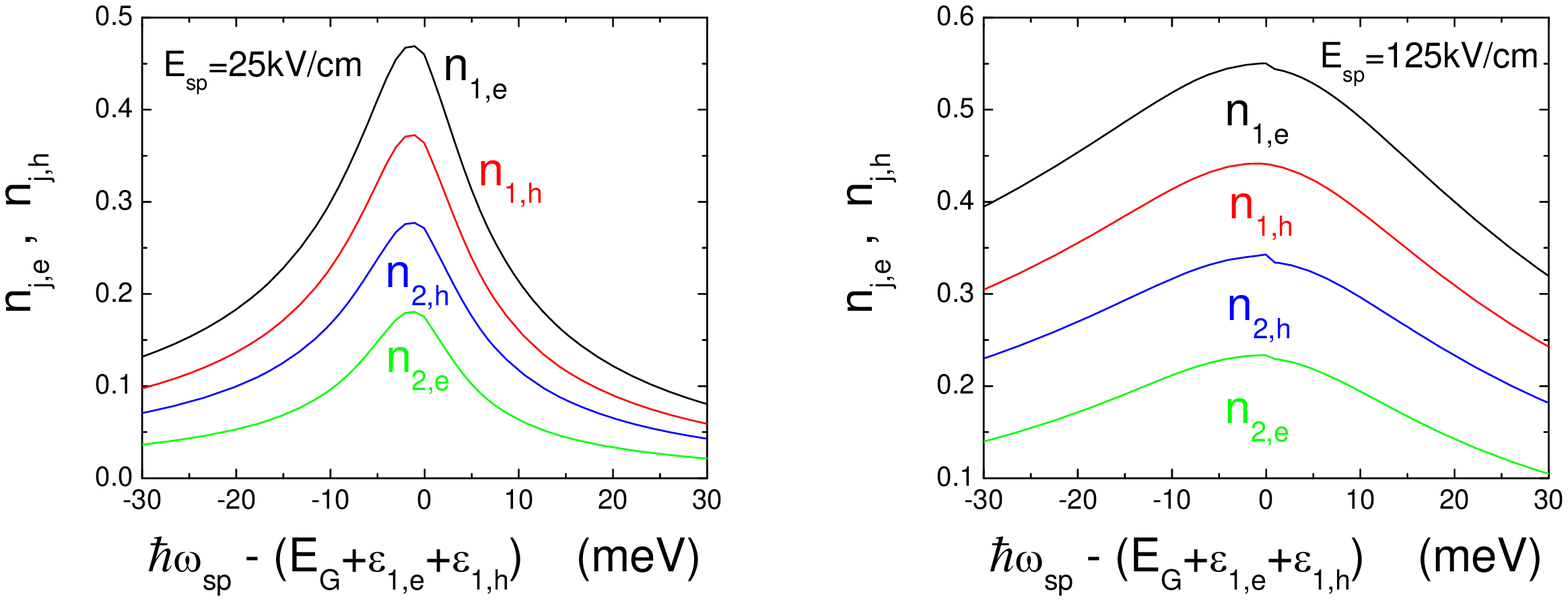,width=0.85\textwidth}
\caption{\label{f3}
(Color online) Optical absorption coefficients $\beta_0(\omega_{\rm sp})$ (upper-left panel) and scattering field $|{\bf E}_{\rm tot}-{\bf E}_{\rm sp}|$ at the quantum dot (upper-right panel),
as well as the energy-level occupations for electrons $n_{\ell,e}$ and holes $n_{j,h}$ (lower panels) with
$\ell,\,j=1,\,2$, as functions of frequency detuning $\Delta\hbar\omega_{\rm sp}\equiv\hbar\omega_{\rm sp}-(E_{\rm G}+\varepsilon_{1,e}+\varepsilon_{1,h})$.
Here, the results for various amplitudes $E_{\rm sp}$ of an SPP field with frequency $\omega_{\rm sp}$ are presented in the upper panels, along with a comparison of the energy-level occupations for
$E_{\rm sp}=25$ and $125$\,kV/cm in the lower panels. The label $\times 0.1$ in the upper-left panel indicates that the result is multiplied by a factor of $0.1$.}
\end{figure}

Figure\ \ref{f3} presents the quantum dot absorption coefficient $\beta_0(\omega_{\rm sp})$ for an SPP field, the scattering field $|{\bf E}_{\rm tot}-{\bf E}_{\rm sp}|$ of the SPP field,
and the energy-level occupations for electrons $n_{\ell,e}$ and holes $n_{j,h}$ with $\ell,\,j=1,\,2$ as functions of frequency detuning
$\Delta\hbar\omega_{\rm sp}\equiv\hbar\omega_{\rm sp}-(E_{\rm G}+\varepsilon_{1,e}+\varepsilon_{1,h})$. A dip is observed at resonance $\Delta\hbar\omega_{\rm sp}=0$
in the upper-left panel, which appears to become deeper with decreasing amplitude $E_{\rm sp}$ of the SPP field in the strong-coupling regime due to a decrease in the saturated absorption.
However, this dip completely disappears when $E_{\rm sp}$ drops to $25$\,kV/cm in the weak-coupling regime due to the suppression of the photon-dressing effect,
which is accompanied by an order of magnitude increase in the absorption-peak strength.
The dip in the upper-left panel corresponds to a peak in the scattering field, as can be seen from the upper-right panel of the figure.
The scattering field increases with frequency detuning away from resonance, corresponding to the decreasing absorption.
As a result, two minima show up on both sides of resonance for the scattering field in the strong-coupling regime.
Maxwell-Bloch equations couple the field dynamics outside of a quantum dot with the electron dynamics inside the dot.
At $E_{\rm sp}=125$\,kV/cm in the lower-right panel, we find peaks in energy-level
occupations at resonance, which are broadened by the finite carrier lifetime as well as the optical power of the SPP field.
Moreover, jumps in the energy-level occupations can be seen at resonance due to Rabi splitting of the energy levels in the dressed electron states.
The effect of resonant phonon absorption also plays a significant role in the finite value of $n_{2,e}$ with energy-level separations $\varepsilon_{2,e}-\varepsilon_{1,e}\approx\hbar\Omega_0$. However, as $E_{\rm sp}$
decreases to $25$\,kV/cm in the lower-left panel, peaks in the energy-level occupations are greatly sharpened and negatively shifted
due to the suppression of the broadening from the optical power and the excitonic effect, respectively.
Additionally, jumps in the energy-level occupations become invisible because the Rabi-split energy gap in this case is much smaller than the energy-level broadening from the finite lifetime of electrons
(i.e. severely damped Rabi oscillations between the first electron and hole levels).
\medskip

\begin{figure}[p]
\centering
\epsfig{file=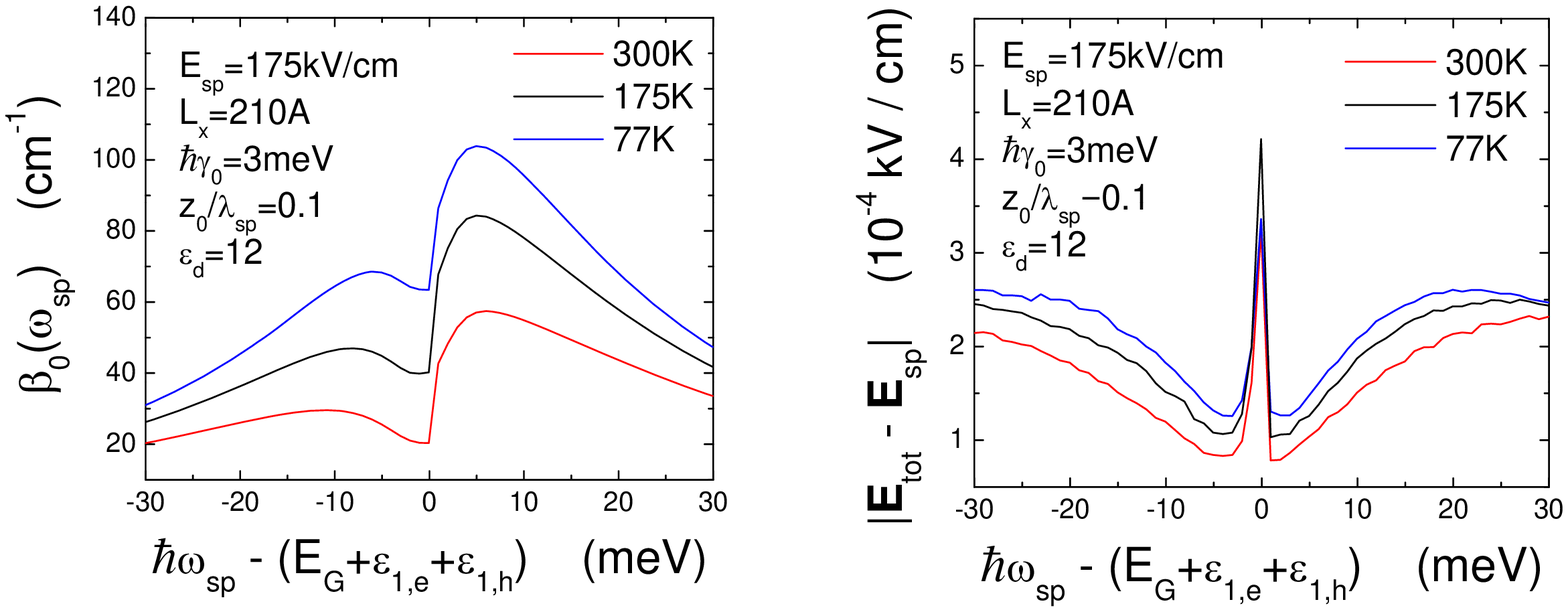,width=0.85\textwidth}
\epsfig{file=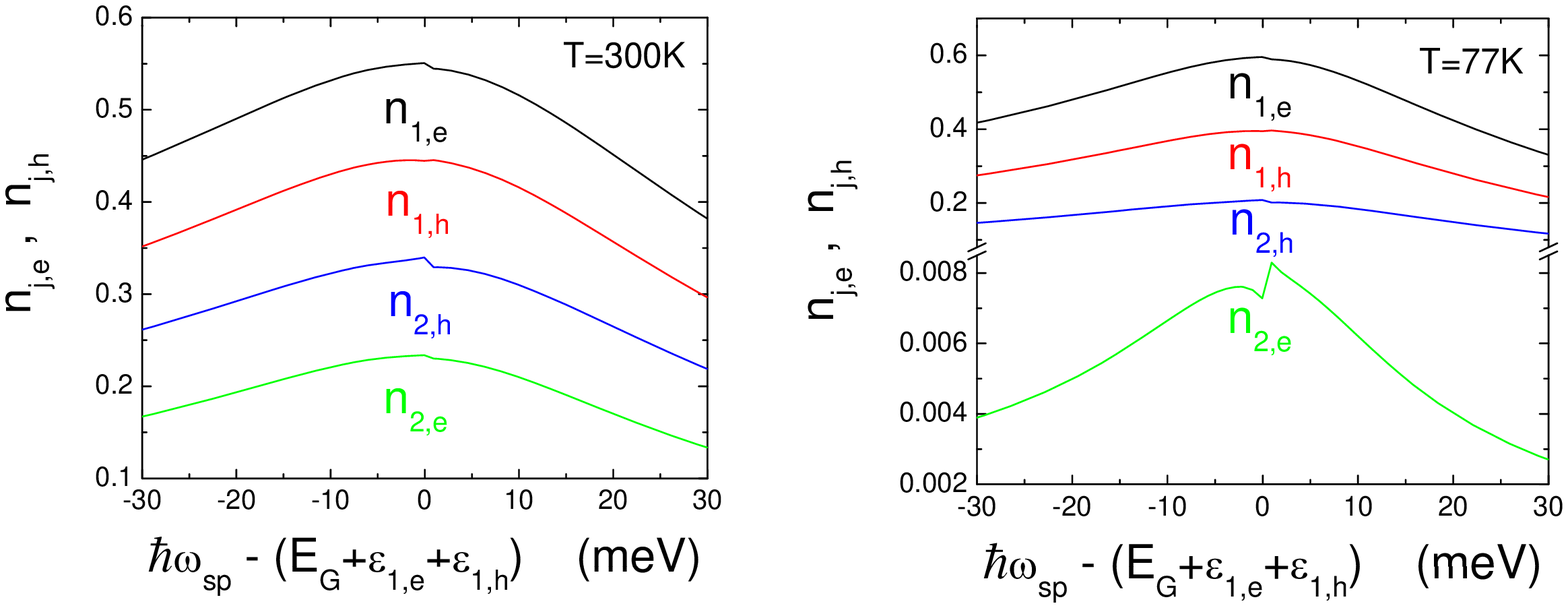,width=0.85\textwidth}
\caption{\label{f4}
(Color online) $\beta_0(\omega_{\rm sp})$ (upper-left panel) and $|{\bf E}_{\rm tot}-{\bf E}_{\rm sp}|$ (upper-right panel),
as well as $n_{\ell,e}$ and holes $n_{j,h}$ (lower panels), as functions of $\Delta\hbar\omega_{\rm sp}$.
Here, the results for three different temperatures $T=300$, $175$ and $77$\,K are displayed in the upper panels, along with a comparison of $n_{\ell,e}$ and holes $n_{j,h}$ for
$T=300$ and $77$\,K in the lower panels.}
\end{figure}

We know that a decrease in temperature $T$ gives rise to an increase in the crystal bandgap $E_{\rm G}$. On the other hand, the localization of an SPP field
(i.e. an exponential decay of the field strength on either side of a metallic surface) is greatly enhanced when the SPP frequency $\omega_{\rm sp}$
approaches that of a surface plasmon. As a result, the field at the quantum dot is expected to decrease as $T$ is reduced.
This gives rise to a higher absorption coefficient for a lower temperature,
as shown in the upper-left panel of Fig.\,\ref{f4}. Interestingly, although the suppressed absorption coefficient can be seen from $\beta_0(\omega_{\rm sp})$ for high SPP-field amplitudes, as shown by Eq.\,(\ref{add2}),
from the upper-right panel of this figure we find the resonant peak at $\hbar\omega_{\rm sp}=E_{\rm G}+\varepsilon_{\rm 1,e}+\varepsilon_{\rm 1,h}$
initially increases with $T$ but then decreases with $T$ at room temperature.
This subtle difference demonstrates the effect of reduced phonon absorption at $T=77$\,K on the resonant scattering field by the factor $1-n_{\rm e}(t)-n_{\rm h}(t)$ in Eq.\,(\ref{e49}).
Moreover, the strong effect of the suppressed optical-phonon absorption between two electron energy levels at $77$\,K
is clearly demonstrated in the lower panels of Fig.\,\ref{f4}, where the level occupation $n_{\rm 2,e}$ becomes negligible at $T=77$\,K in comparison with that at $T=300$\,K.
\medskip

\begin{figure}[p]
\centering
\epsfig{file=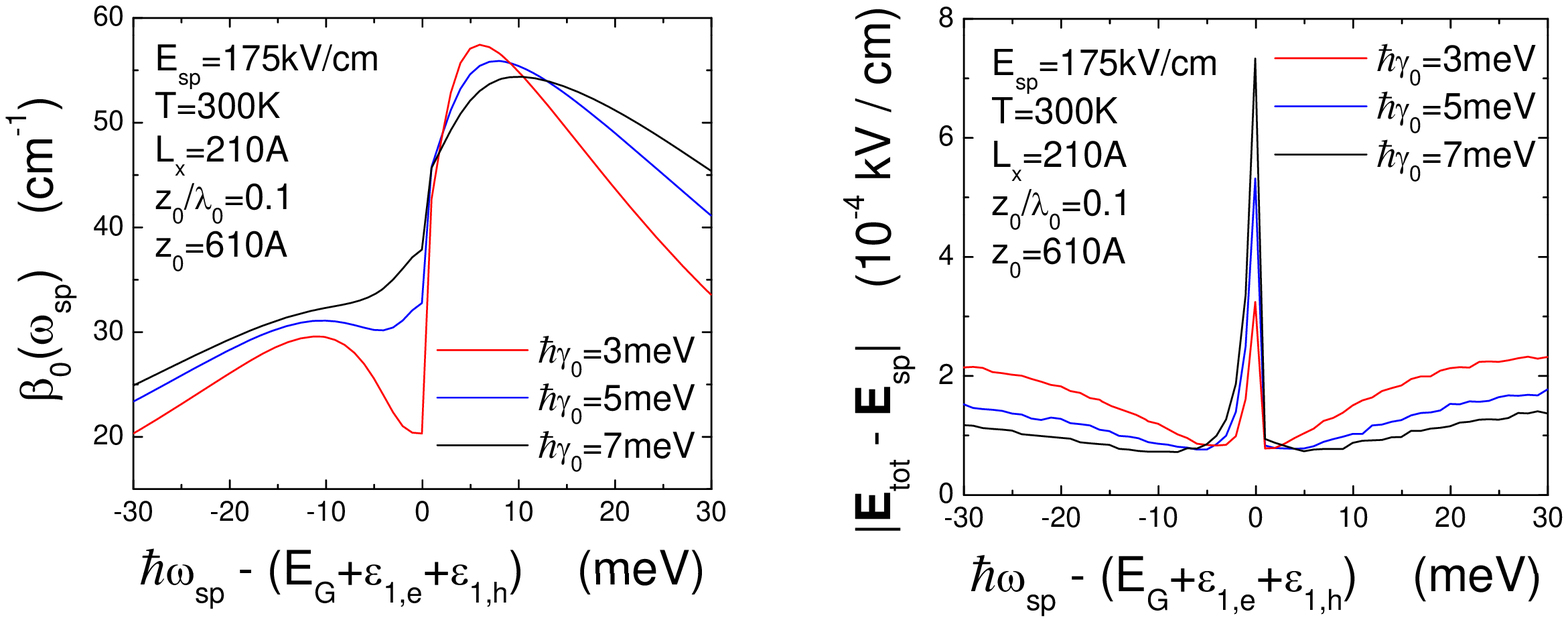,width=0.85\textwidth}
\epsfig{file=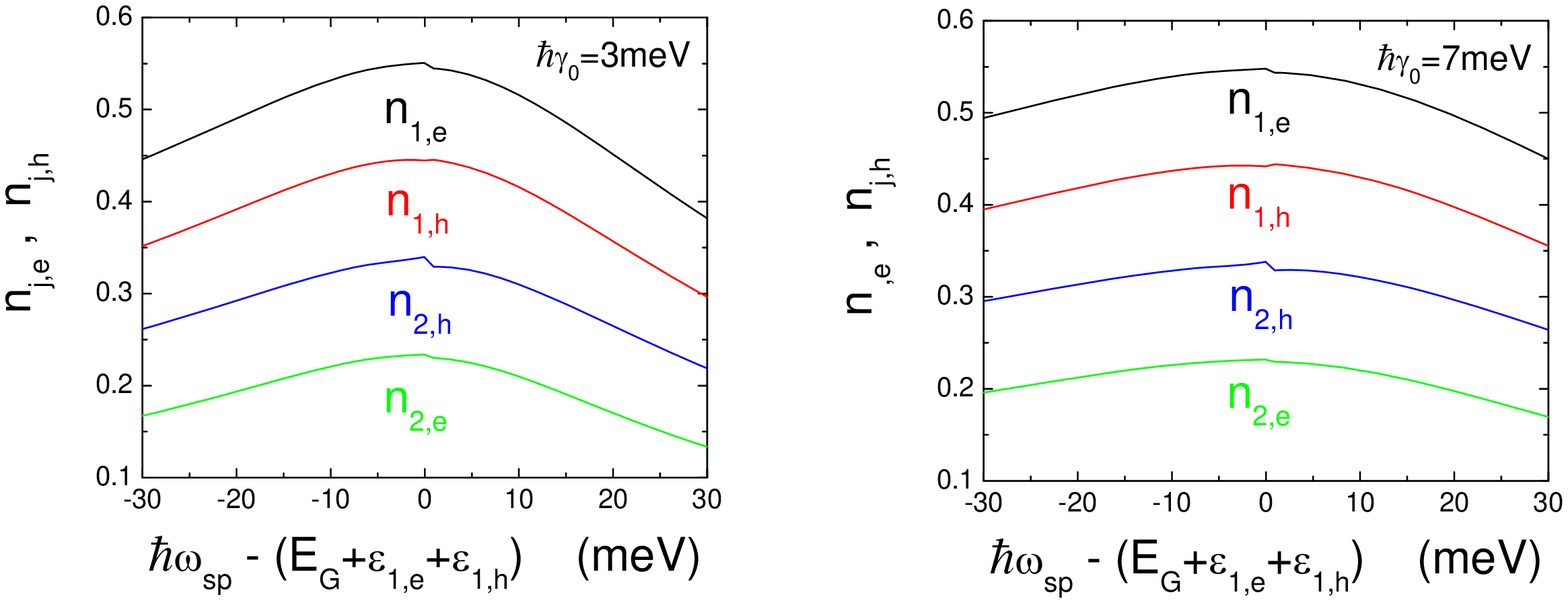,width=0.85\textwidth}
\caption{\label{f5}
(Color online) $\beta_0(\omega_{\rm sp})$ (upper-left panel) and $|{\bf E}_{\rm tot}-{\bf E}_{\rm sp}|$ (upper-right panel),
as well as $n_{\ell,e}$ and holes $n_{j,h}$ (lower panels), as functions of $\Delta\hbar\omega_{\rm sp}$.
Here, the results with different energy-level broadening $\hbar\gamma_0=3$, $5$ and $7$\,meV are shown in the upper panels, along with a comparison of $n_{\ell,e}$ and holes $n_{j,h}$ for
$\hbar\gamma_0=3$ and $7$\,meV in the lower panels.}
\end{figure}

The electron thermal dynamics due to phonon absorption has been demonstrated in Fig.\,\ref{f4} for various temperatures. In Fig.\,\ref{f5}, we present the electron dynamics resulting from the optical dephasing,
due to the finite lifetime of electrons, at different energy-level broadenings $\hbar\gamma_0$. As $\hbar\gamma_0$ is increased from $3$\,meV to $7$\,meV, the dip in $\beta_0(\omega_{\rm sp})$ at resonance is suppressed,
leading to a single peak with a reduced strength and an increased width, as shown in the upper-left panel of the figure. This increase in the resonant absorption
is further accompanied by an enhanced resonant peak for the scattering field in the upper-right
panel of this figure. As expected, the energy-level occupations at $\hbar\gamma_0=7$\,meV become much broader than those at $\hbar\gamma_0=3$\,meV, as displayed in the lower two panels of the figure.
\medskip

\begin{figure}[p]
\centering
\epsfig{file=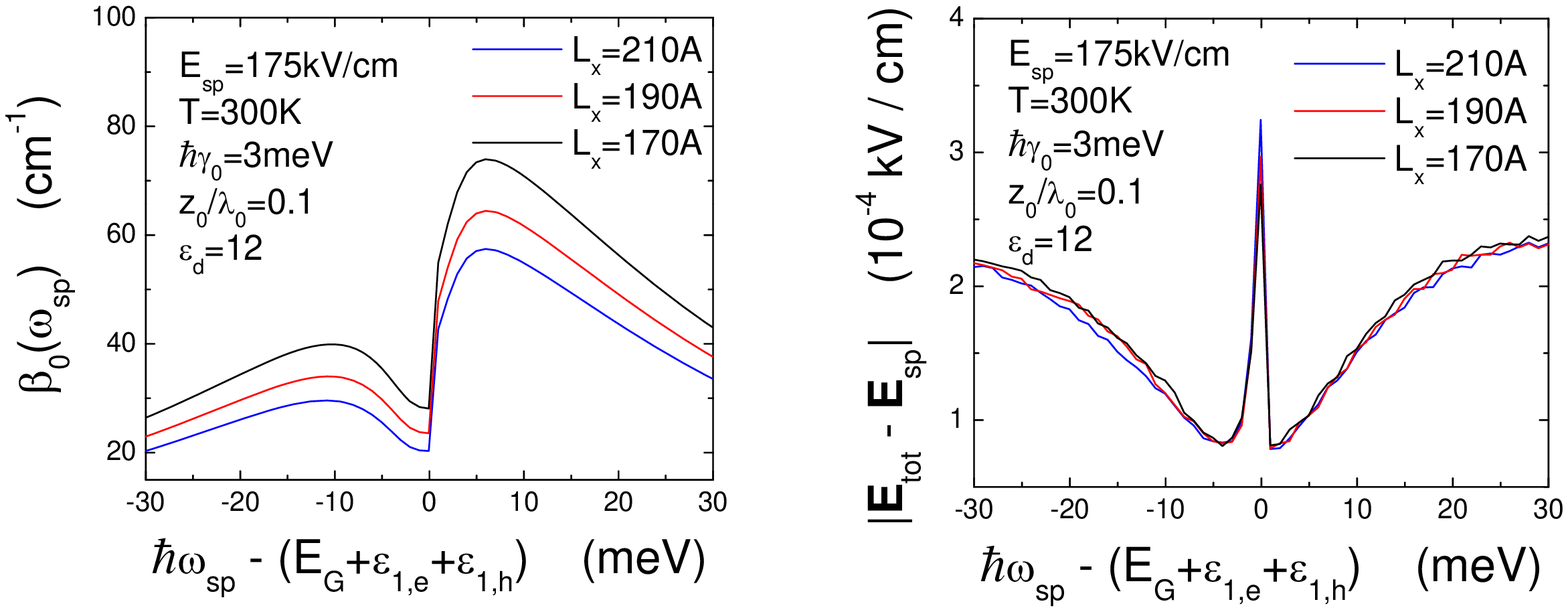,width=0.85\textwidth}
\epsfig{file=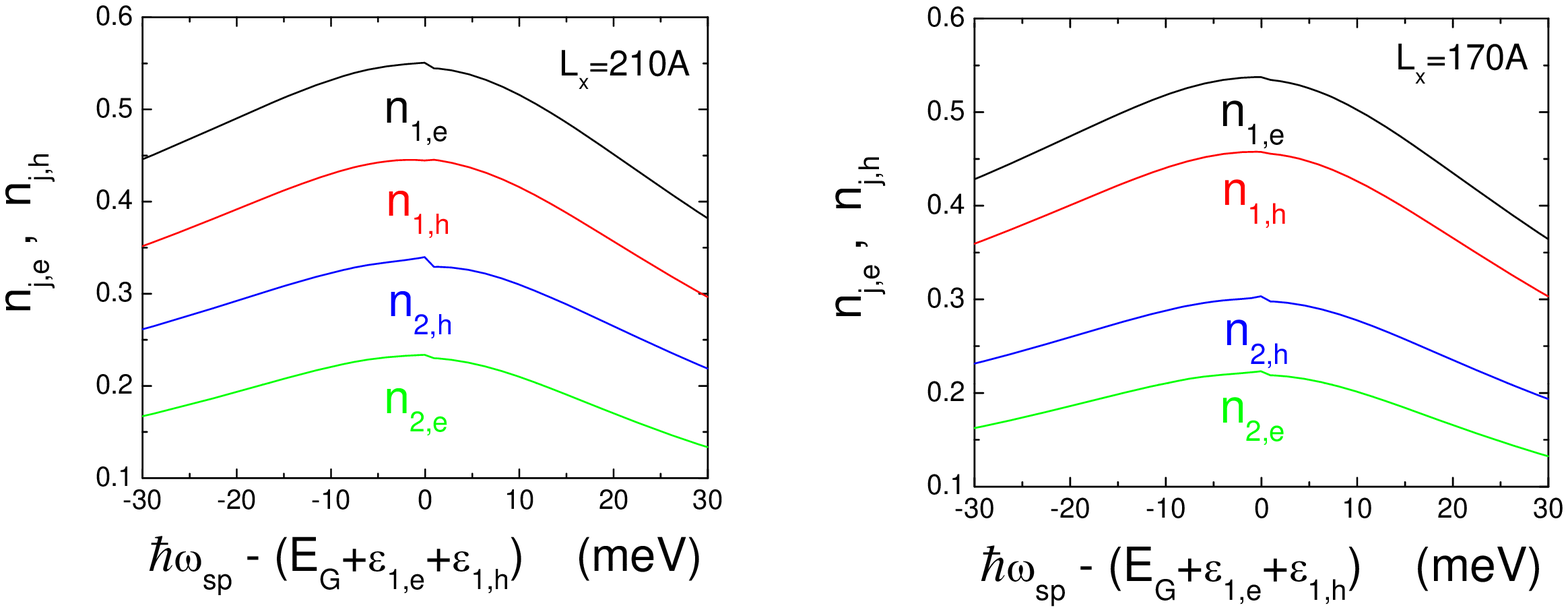,width=0.85\textwidth}
\caption{\label{f6}
(Color online) $\beta_0(\omega_{\rm sp})$ (upper-left panel) and $|{\bf E}_{\rm tot}-{\bf E}_{\rm sp}|$ (upper-right panel),
as well as $n_{\ell,e}$ and holes $n_{j,h}$ (lower panels), as functions of $\Delta\hbar\omega_{\rm sp}$.
Here, the results for three different sizes $L_x=210$, $190$ and $170$\,\AA\ of a quantum dot are shown in the upper panels, along with a comparison of $n_{\ell,e}$ and holes $n_{j,h}$ for
$L_x=210$ and $170$\,\AA\ in the lower panels.}
\end{figure}

We further notice that the effective bandgap $E_{\rm G}+\varepsilon_{\rm 1,e}+\varepsilon_{\rm 1,h}$ also
depends on the size $L_x$ of a quantum dot due to the quantization effect, and the effective bandgap will increase with decreasing $L_x$. The size effect from such an $L_x$ dependence is displayed in Fig.\,\ref{f6}.
From the upper-left panel of Fig.\,\ref{f6}, we find that the peak of $\beta_0(\omega_{\rm sp})$
is enhanced as $L_x$ is reduced. This phenomenon is connected to the increased localization of the SPP field at $L_x=170$\,\AA as the SPP frequency approaches the saturation part of its dispersion.
Moreover, the dip in $\beta_0(\omega_{\rm sp})$ is lifted somewhat uniformly at the same time due to decreased $n_1^{\rm e}(t)$ from the enhanced Coulomb and phonon scattering at $L_x=170$\,\AA.
Here, $\beta_0(\omega_{\rm sp})$ is proportional to the population factor $1-n^{\rm e}_1(t)-n_1^{\rm h}(t)$, as can be seen from Eq.\,(\ref{add2}).
Besides the slightly-reduced resonant peak strength of the scattering field for $L_x=170$\,\AA\ (also resulting from the enhanced carrier scattering),
$|{\bf E}_{\rm tot}-{\bf E}_{\rm sp}|$ keeps the same peak position, as shown in the upper-right panel of the figure.
In this case, $|{\bf E}_{\rm tot}-{\bf E}_{\rm sp}|$ at the dot approaches a nonzero value at resonance, as can be seen from
Eq.\,(\ref{e55}), and tends to zero rapidly away from resonance.
Additionally, $n_{\rm 2,h}$ is reduced for $L_x=170$\,\AA, as can be found from a comparison between the two lower panels of the figure. This is attributed to the reduced phonon absorption between two hole energy levels.
\medskip

\begin{figure}[p]
\centering
\epsfig{file=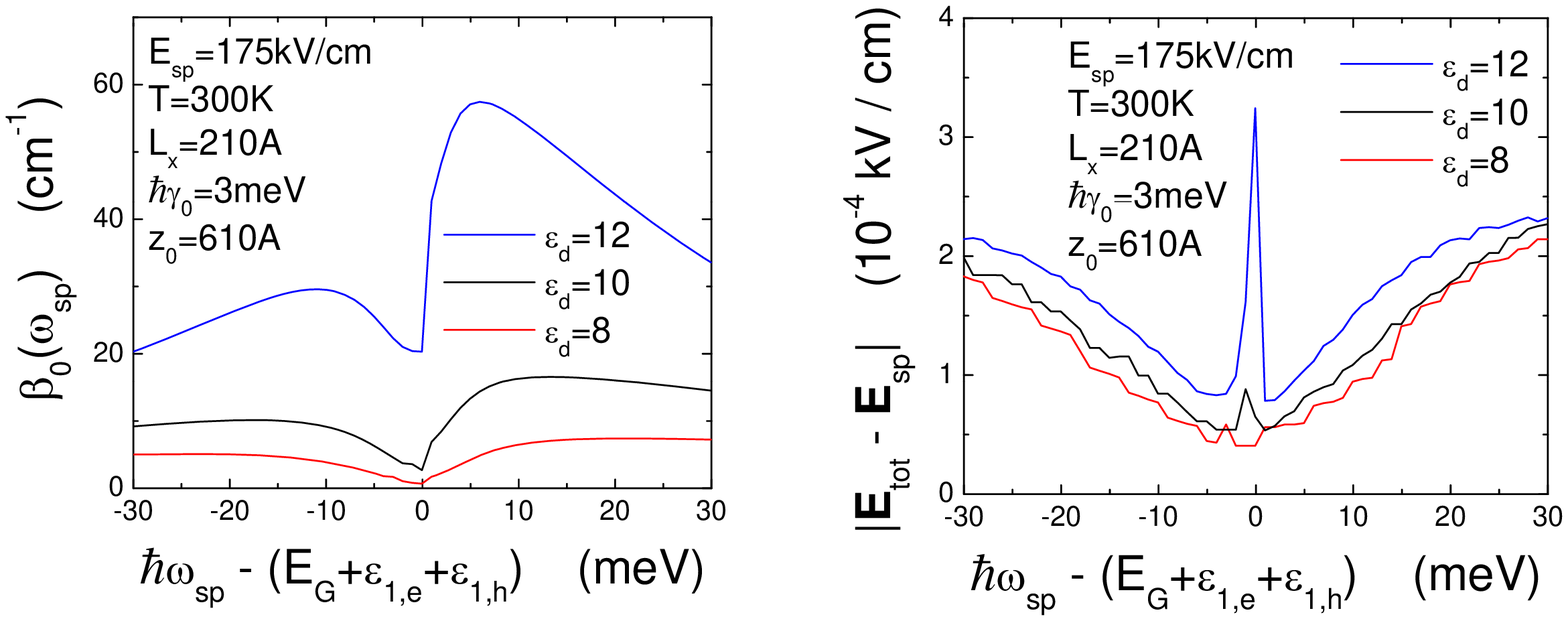,width=0.85\textwidth}
\epsfig{file=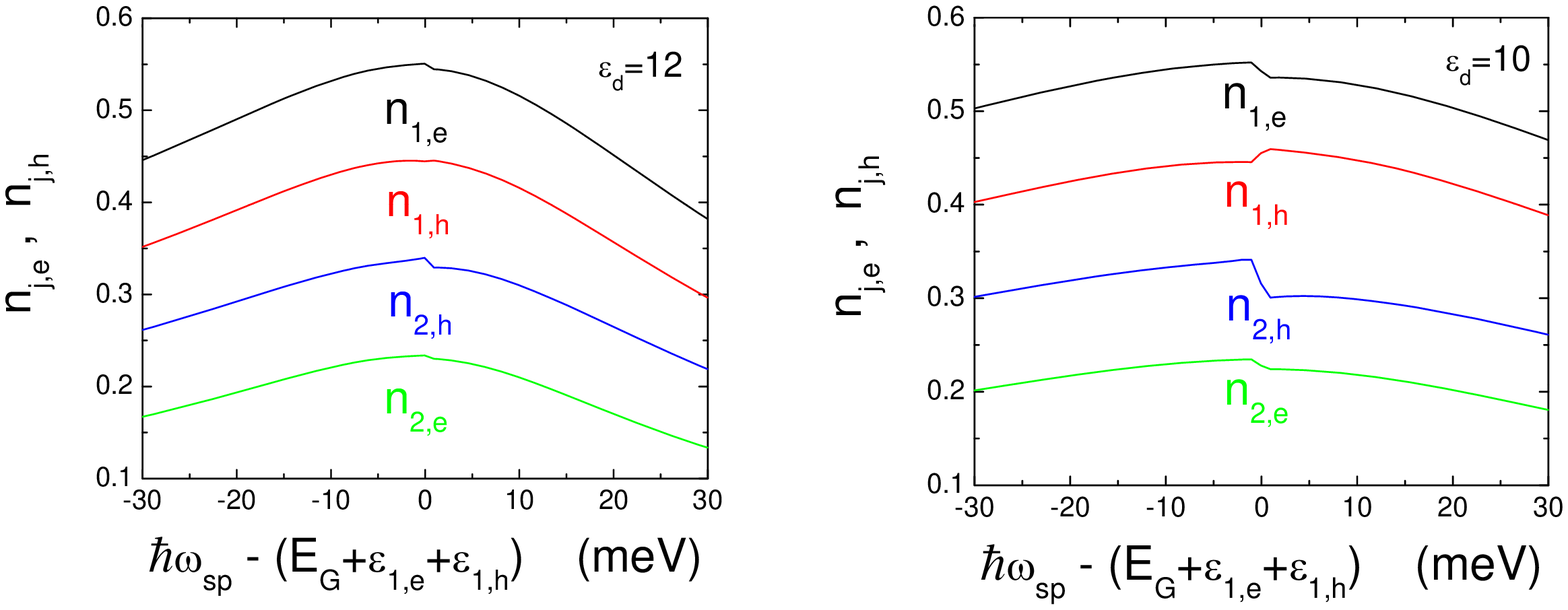,width=0.85\textwidth}
\caption{\label{f7}
(Color online) $\beta_0(\omega_{\rm sp})$ (upper-left panel) and $|{\bf E}_{\rm tot}-{\bf E}_{\rm sp}|$ (upper-right panel),
as well as $n_{\ell,e}$ and holes $n_{j,h}$ (lower panels), as functions of $\Delta\hbar\omega_{\rm sp}$.
Here, the results for three dielectric constants of a cladding layer, with $\epsilon_{\rm d}=8$, $10$ and $12$, are displayed in the upper panels, along with a comparison of $n_{\ell,e}$ and holes $n_{j,h}$ for
$\epsilon_{\rm d}=12$ and $10$ in the lower panels.}
\end{figure}

In Figs.\,\ref{f4} and \ref{f6}, we vary the localization of an SPP field by changing the effective bandgap. Since the frequency of the surface plasmon (saturated dispersion part) is proportional to
the factor of $1/\sqrt{1+\epsilon_{\rm d}}$, a smaller value of $\epsilon_{\rm d}$ implies a higher surface-plasmon frequency or a reduced localization of the SPP field. We verify the change in the SPP localization
by observing the upper two panels of Fig.\,\ref{f7}, where the absorption peak, as well as the resonant scattering-field peak, become much stronger as $\epsilon_{\rm d}$ is increased from $8$ to $12$
due to the reduction of saturated absorption for a lower field strength at the quantum dot. Furthermore, from the two lower panels of this figure
we also observe, via the jumps in the population curves, an enhanced Rabi-split energy gap in the electron dressed states as $\epsilon_{\rm d}$ is reduced from $12$ to $10$ due to the enhanced field strength at the quantum dot.
\medskip

\begin{figure}[p]
\centering
\epsfig{file=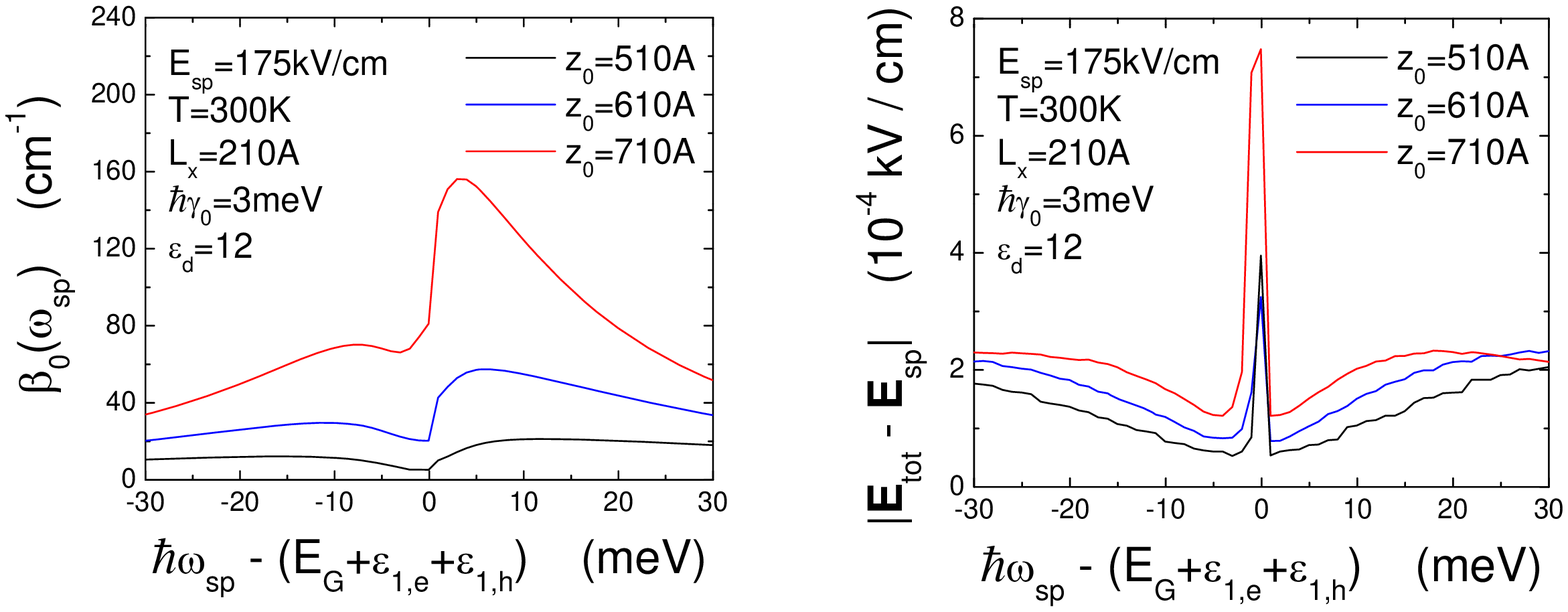,width=0.85\textwidth}
\epsfig{file=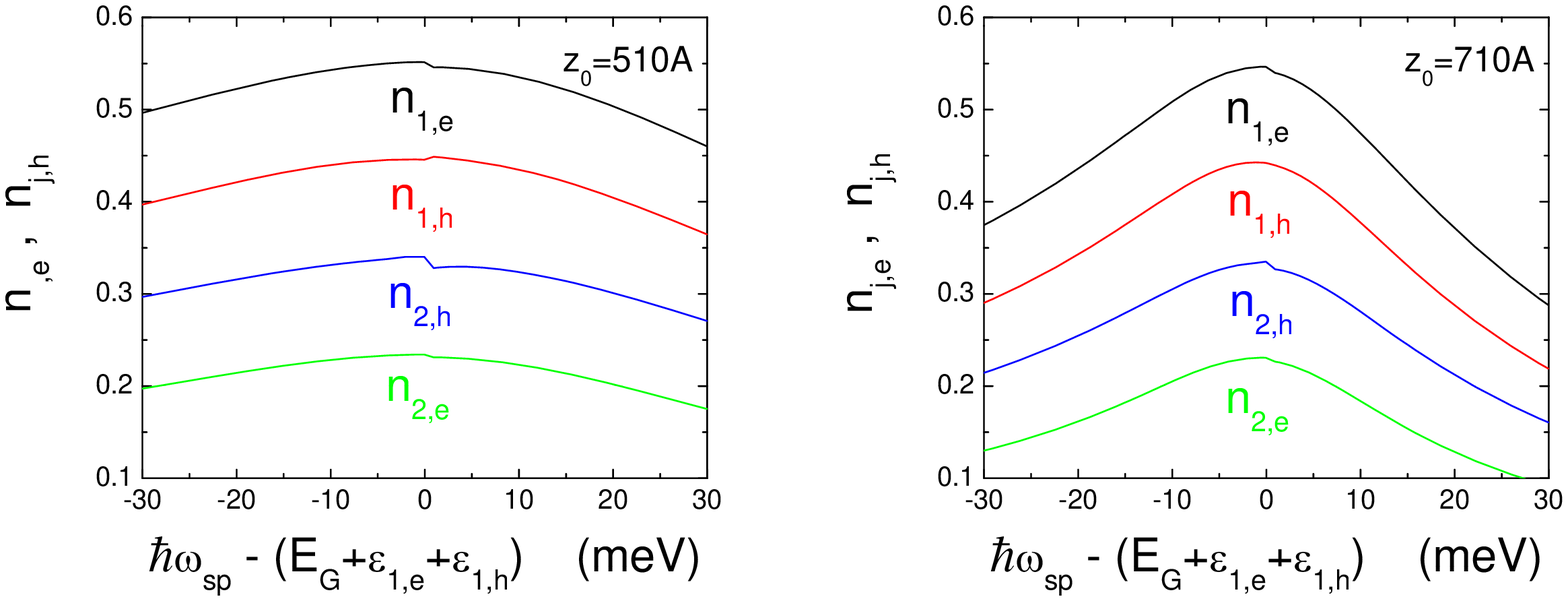,width=0.85\textwidth}
\caption{\label{f8}
(Color online) $\beta_0(\omega_{\rm sp})$ (upper-left panel) and $|{\bf E}_{\rm tot}-{\bf E}_{\rm sp}|$ (upper-right panel),
as well as $n_{\ell,e}$ and holes $n_{j,h}$ (lower panels), as functions of $\Delta\hbar\omega_{\rm sp}$.
Here, the results for different vertical distances of a quantum dot from a metallic surface are displayed in the upper panels for $z_0=510$, $610$ and $710$\,\AA,
along with a comparison of $n_{\ell,e}$ and holes $n_{j,h}$ for $z_0=510$ and $710$\,\AA\ in the lower panels.}
\end{figure}

In the presence of the localization of an SPP field, we can move a quantum dot closer to a metallic surface to gain a higher field at the quantum dot.
The upper-left panel of Fig.\,\ref{f8} has elucidated this fact,
in which a larger $z_0$ corresponds to a weaker field, and then, a higher absorption peak due to the reduction of saturated absorption.
This fact is also reflected in the upper-right panel of the figure, where a higher
resonant scattering-field peak occurs for a larger value of $z_0$. At $z_0=510$\,\AA, a Rabi-split energy gap at resonance is clearly visible from the lower-left-panel of the figure for electron dressed states.
Additionally, at $z_0=710$\,\AA, by entering into a weak-coupling regime for a weaker field at the dot, we find sharpened resonant peaks in the energy-level occupations of electrons and holes,
similar to the observation from the lower-left panel of Fig.\,\ref{f3}.

\subsection{Results for the dressed states of electrons}

In the second part of the numerical calculations, besides the parameters given in the first subsection, we have fixed $L_x=210$\,\AA, $\hbar\gamma_0=3$\,meV, $z_0=610$\,\AA\ and $\epsilon_{\rm d}=12$.
Other parameters, including $T$, $E_{\rm sp}$ and $\Delta\hbar\omega_{\rm sp}$, will be directly indicated in the figures. Additionally, $\Delta\hbar\omega_{\rm sp}$ is given with respect to the energy gap at $T=300$\,K.
\medskip

\begin{figure}[p]
\centering
\epsfig{file=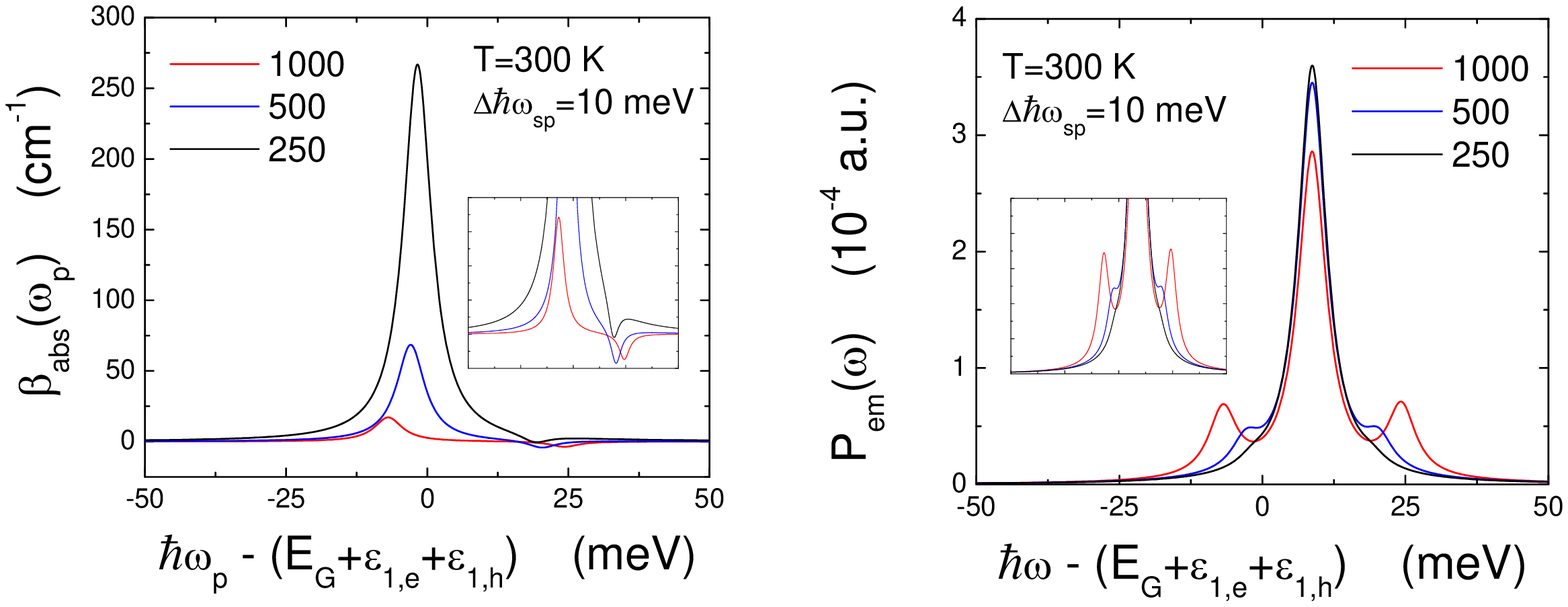,width=0.9\textwidth}
\caption{\label{f9}
(Color online) Probe-field absorption coefficient $\beta_{\rm abs}(\omega_{\rm p})$ (left panel) and spontaneous emission of a quantum dot $P_{\rm em}(\omega)$ (right panel)
as a function of $\hbar\omega_{\rm p}-(E_{\rm G}+\varepsilon_{\rm 1,e}+\varepsilon_{\rm 1,h})$ and $\hbar\omega-(E_{\rm G}+\varepsilon_{\rm 1,e}+\varepsilon_{\rm 1,h})$, respectively, are presented.
Here, comparisons of the results with three values of SPP-field amplitudes $E_{\rm sp}=250$, $500$ and $1000$\,kV/cm are given. The insets of both panels are enlarged views of the peaks.}
\end{figure}

From the left panel of Fig.\,\ref{f9} we find a strong absorption (positive) peak and a weak gain (negative) peak for the probe-field absorption coefficient $\beta_{\rm abs}(\omega_{\rm p})$
due to a quantum coherence effect from the electron states being dressed by an SPP field. In the strong-coupling regime, the dispersion of the quantum-dot e-h plasmas (dot-like branch)
and SPPs (photon-like branch) form an anticrossing gap, where a higher-energy dot-like branch at a negative frequency detuning switches to a photon-like branch for a positive detuning.
The positive peak is associated with the absorption of a probe-field photon by a quantum-dot e-h plasma,
while the negative peak relates to the process with absorption of two photons from an SPP field and emission of one probe-field photon. The absorption peak is significantly reduced by saturation
at $E_{\rm sp}=1000$\,kV/cm, and the gain peak is suppressed by a smaller Rabi-coupling frequency at  $E_{\rm sp}=250$\,kV/cm (see the inset of the left panel).
In addition, we observe from the right panel of Fig.\,\ref{f9} that
two Rabi-splitting-induced side emission peaks for the spontaneous emission $P_{\rm em}(\omega)$ become weaker and closer to the strong central peak as $E_{\rm sp}$ is reduced (see the inset of the right panel).
Moreover, the strength of the central peak due to the coherent conversion of an absorbed SPP-field photon to a
spontaneously-emitted photon  (non-linear optical behavior) is slightly reduced at $E_{\rm sp}=1000$\,kV/cm as a result of saturated absorption of the SPP field.
\medskip

\begin{figure}[p]
\centering
\epsfig{file=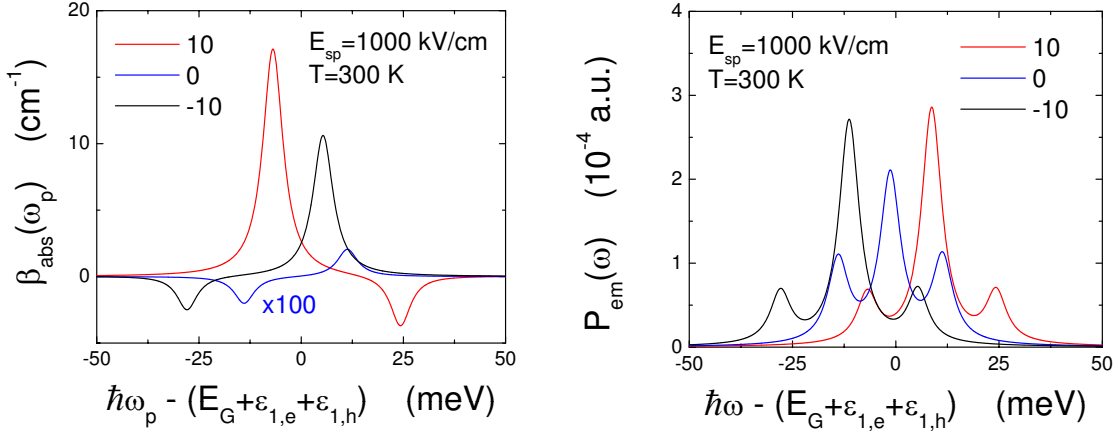,width=0.9\textwidth}
\caption{\label{f10}
(Color online) $\beta_{\rm abs}(\omega_{\rm p})$ (left panel) and $P_{\rm em}(\omega)$ (right panel)
as a function of $\hbar\omega_{\rm p}-(E_{\rm G}+\varepsilon_{\rm 1,e}+\varepsilon_{\rm 1,h})$ and $\hbar\omega-(E_{\rm G}+\varepsilon_{\rm 1,e}+\varepsilon_{\rm 1,h})$, respectively, are displayed.
Here, we show the comparisons with three values of SPP-field detunings $\Delta\hbar\omega_{\rm sp}=0$ and $\pm 10$\,meV. The label $\times 100$ for zero SPP-field detuning
indicates the result is multiplied by a factor of $100$.}
\end{figure}

Figure\ \ref{f10} demonstrates the effect of frequency detuning $\Delta\hbar\omega_{\rm sp}$ of an SPP field with respect to the bandgap of a quantum dot.
The switching of the detuning from $10$\,meV to $-10$\,meV reveals the corresponding spectral-position interchange between the absorption (dot-like branch) and the gain (photon-like branch)
peaks for $\beta_{\rm abs}(\omega_{\rm p})$ in the left panel of the figure.
The Rabi oscillations between the first electron and hole energy levels are weakened with increasing $|\Delta\hbar\omega_{\rm sp}|$.
At resonance with a zero detuning, both the absorption and gain peaks are suppressed by very strong Rabi oscillations.
This detuning also shifts the emission peaks correspondingly because of the coherent conversion of an SPP-field photon to a spontaneously-emitted one, as can been seen from the right panel of this figure.
Moreover, the central peak is weakened and the two side peaks are enlarged at resonance as a result of energy transfer to the side peaks by strong coupling and enhanced Rabi oscillations, respectively.
\medskip

\begin{figure}[p]
\centering
\epsfig{file=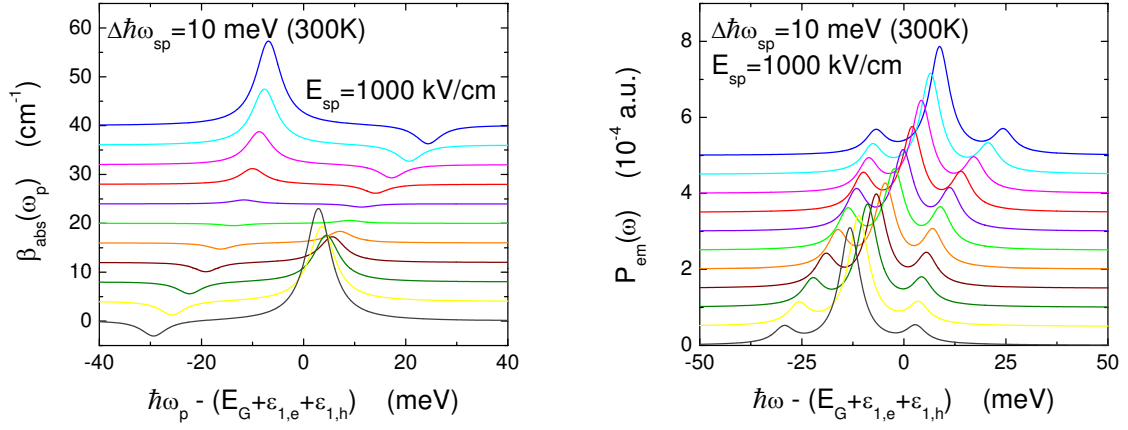,width=0.9\textwidth}
\caption{\label{f11}
(Color online) $\beta_{\rm abs}(\omega_{\rm p})$ (left panel) and $P_{\rm em}(\omega)$ (right panel)
as a function of $\hbar\omega_{\rm p}-(E_{\rm G}+\varepsilon_{\rm 1,e}+\varepsilon_{\rm 1,h})$ and $\hbar\omega-(E_{\rm G}+\varepsilon_{\rm 1,e}+\varepsilon_{\rm 1,h})$, respectively, are shown.
Here, a series of offset curves are displayed for various temperatures from $T=250$\,K (bottom black curves) to $T=300$\,K (top blue curves) in steps of $5$\,K.}
\end{figure}

Since the temperature affects the crystal bandgap energy $E_{\rm G}$, by changing the temperature we are able to scan the detuning $\Delta\hbar\omega_{\rm sp}$ of the SPP field with a fixed SPP frequency
$\hbar\omega_{\rm sp}$ from negative to positive or vice versa. This leads to a spectral-position interchange between the absorption and gain peaks, similar to Fig.\,\ref{f10}.
The results in Fig.\,\ref{f11} prove such an expected feature by increasing $T$ from $250$ to $300$\,K in steps of $5$\,K.
Technically, changing the temperature in the experiment is much easier than changing the tuning of a laser frequency over a large range. Here, the shift of the central peak in the right panel of the
figure directly reflects the variation of the SPP-field detuning with $T$. Furthermore, the interchange between the dot-like and photon-like modes with $T$ in the left panel can be regarded as direct evidence
for the existence of an anticrossing energy gap resulting from a strongly-coupled e-h plasma and SPP field or coupled e-h plasmas and surface plasmons.

\subsection{Time-resolved optical spectra}

\begin{figure}[p]
\centering
\epsfig{file=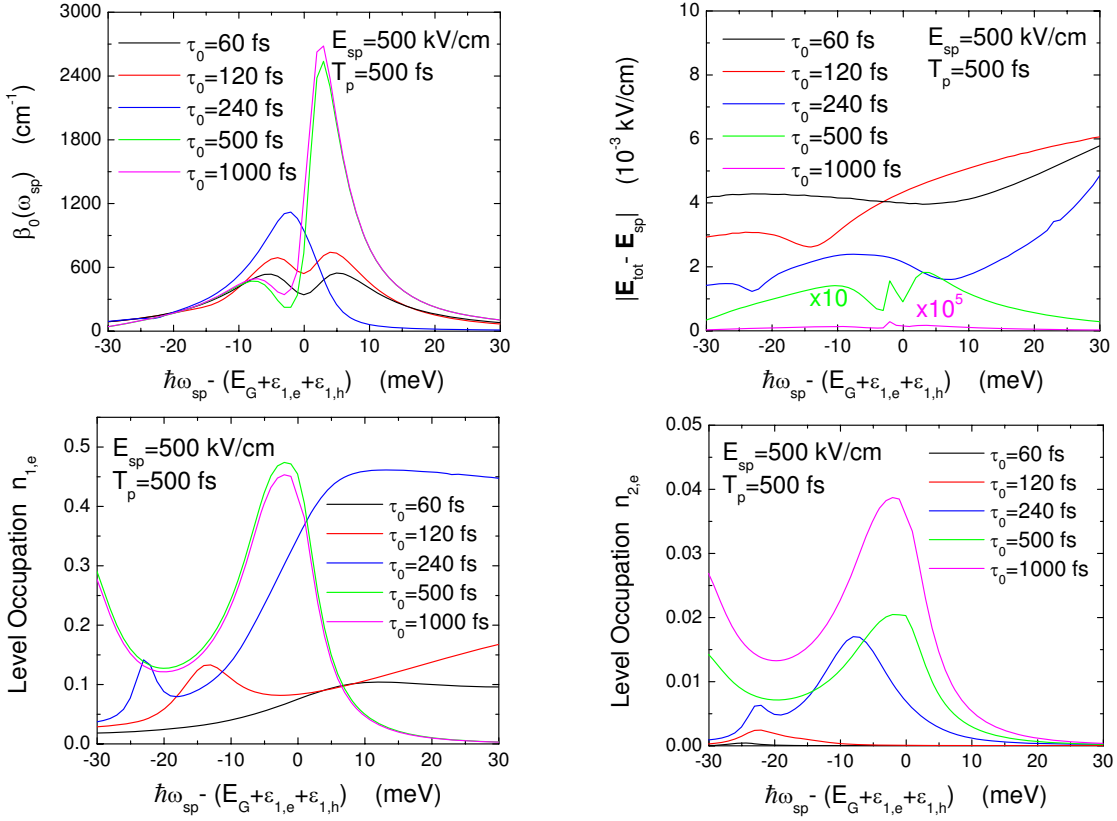,width=0.9\textwidth}
\caption{\label{f12}
(Color online) $\beta_0(\omega_{\rm sp})$ (upper-left panel) and $|{\bf E}_{\rm tot}-{\bf E}_{\rm sp}|$ (upper-right panel),
as well as $n_{1,e}$ and holes $n_{2,e}$ (lower panels), as functions of $\Delta\hbar\omega_{\rm sp}$ at different detection times $\tau_0$ for a Gaussian-shape laser pulse with pulse width $T_p=500$\,fs.
Here, $E_{\rm sp}=500$\,kV/cm is taken, and the other parameters are the same as those in Fig.\,\ref{f3}.
The labels $\times 10$ and $\times 10^5$ in the upper-right panel indicate that the results are multiplied by factors of $10$ and $10^5$, respectively.}
\end{figure}

In our previously presented numerical results, we only showed steady-state dynamics of photo-excited e-h plasmas in a quantum dot
by using a continuous SPP field, where the effects of both phonon scattering and e-h pair
radiative recombination are combined with each other.
Using a laser pulse to launch a pulsed SPP field, we are able to study the dynamics of phonon scattering (narrow pulse) as well as the dynamics of e-h pair radiative recombination (wide pulse), separately.
Dynamically, phonon scattering becomes effective only after a characteristic time (around $1$\,ps), its effect can be seen from a significant increase of $n_{\rm 2,e}$ in our system.
Figure\ \ref{f12} displays the results for
$\beta_0(\omega_{\rm sp})$ (upper-left), $|{\bf E}_{\rm tot}-{\bf E}_{\rm sp}|$ (upper-right), $n_{\rm 1,e}$ (lower-left) and $n_{\rm 2,e}$ (lower-right) for various detection times $\tau_0$
in the presence of a narrow laser pulse (with pulse width $T_p=500$\,fs and peak value $E_{\rm sp}=500$kV/cm) which is turned on at $t=0$. We see from Fig.\,\ref{f12} that $\beta_0(\omega_{\rm sp})$ starts with
a dip for the dressed state at resonance, then shifts to a single peak (at half-pulse width) due to a suppression of the photon-dressing effect. It eventually becomes a single peak plus a shifted dip
after the pulse has passed due to formation of resonant peaks in $n_{\rm 1,e}$ and $n_{\rm 1,h}$.
Correspondingly, $|{\bf E}_{\rm tot}-{\bf E}_{\rm sp}|$ starts by showing a non-resonant behavior with a relatively large magnitude, then shifts to a quasi-resonant behavior, and finally looks like suppressed
resonant behavior with a peak at and dips on both sides of $\Delta\hbar\omega_{\rm sp}=0$. The resonant build up of $n_{\rm 1,e}$ after $\tau_0\geq 500$\,fs can also be verified from this figure, which is
accompanied by the start of significant phonon absorption after $\tau_0\geq 1$\,ps.
\medskip

\begin{figure}[p]
\centering
\epsfig{file=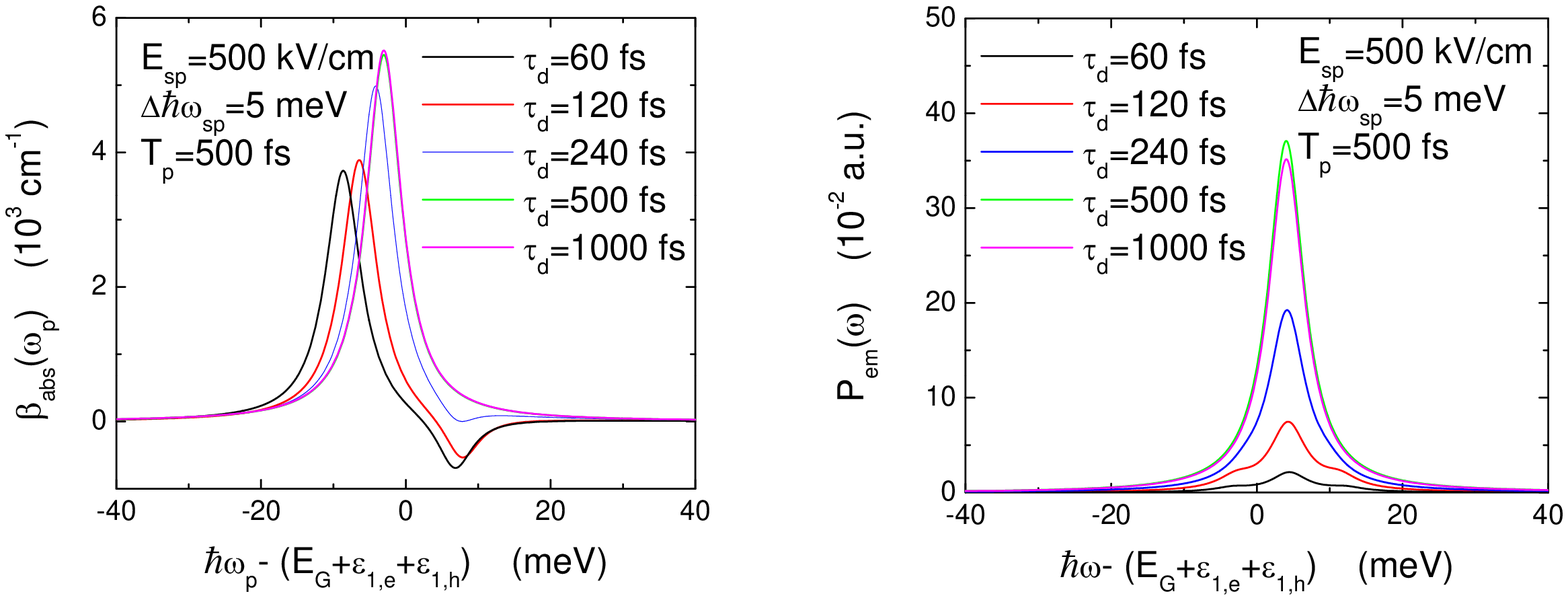,width=0.9\textwidth}
\caption{\label{f13}
(Color online) $\beta_{\rm abs}(\omega_{\rm p})$ (left panel) and $P_{\rm em}(\omega)$ (right panel)
as a function of $\hbar\omega_{\rm p}-(E_{\rm G}+\varepsilon_{\rm 1,e}+\varepsilon_{\rm 1,h})$ and $\hbar\omega-(E_{\rm G}+\varepsilon_{\rm 1,e}+\varepsilon_{\rm 1,h})$, respectively,
at different delayed times $\tau_d$ for a Gaussian-shape laser pulse with pulse width $T_p=500$\,fs.
Here, $E_{\rm sp}=500$\,kV/cm and $\Delta\hbar\omega_{\rm sp}=5$\,meV are chosen, and the other parameters are the same as those in Fig.\,\ref{f3}.}
\end{figure}

Technically, detecting dynamics of photo-excited e-h plasmas by using another time-delayed weak probe field is much more feasible, as shown in Fig.\,\ref{f13}.
From the left panel of this figure, we find that $\beta_{\rm abs}(\omega_{\rm p})$ starts with a pair of positive absorption and negative gain peaks due to a very strong photon dressing effect
for the delayed times $\tau_d=60$ and $120$\,fs. This is changed to a strong absorption peak plus a very weak gain peak at $\tau_d=240$\,fs. At the end, $\beta_{\rm abs}(\omega_{\rm p})$ becomes
independent of $\tau_d$, indicating that a linear optical-response regime has been reached. On the other hand, from the right panel of this figure, we see that the central peak of $P_{\rm em}(\omega)$ is
gradually built up with increasing $\tau_d$ due to enhanced $n_{\rm 1,e}$ and $n_{\rm 1,h}$ around resonance, while two side peaks become weakened and disappear at the same time due to weakened Rabi oscillations.
Interestingly, we also find that the central peak of $P_{\rm em}(\omega)$ slightly decreases at $\tau_d=1$\,ps, which agrees with the observed start of significant phonon absorption seen in
the lower-left panel of Fig.\,\ref{f12}.
\medskip

\begin{figure}[p]
\centering
\epsfig{file=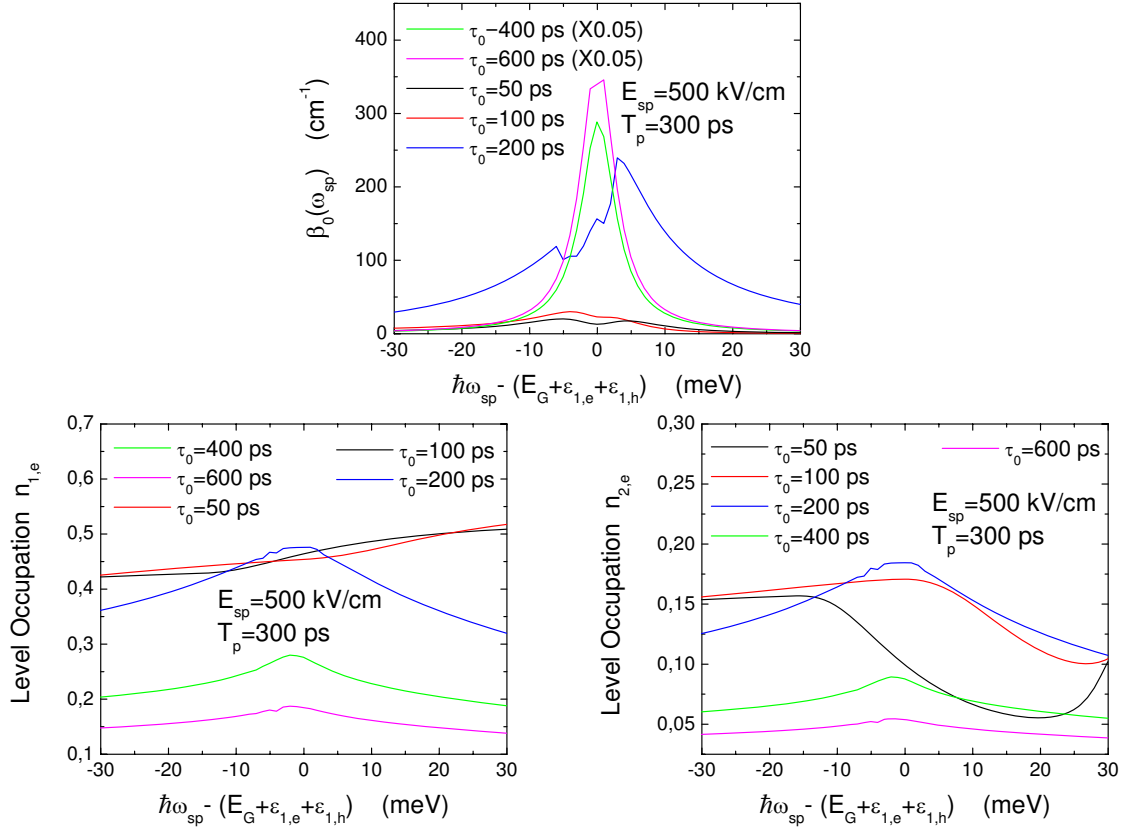,width=0.9\textwidth}
\caption{\label{f14}
(Color online) $\beta_0(\omega_{\rm sp})$ (upper-middle panel), $n_{1,e}$ and holes $n_{2,e}$ (lower two panels), as functions of $\Delta\hbar\omega_{\rm sp}$ at different detection times $\tau_0$ for a wide
laser pulse with pulse width $T_p=300$\,ps.
Here, $E_{\rm sp}=500$\,kV/cm is taken, and the other parameters are the same as those in Fig.\,\ref{f3}.
The labels $\times 0.05$ in the upper-middle panel indicate that the results in the upper panel for $\tau_0=60$ and $120$\,ps are multiplied by a factor of $0.05$.}
\end{figure}

In order to explore the dynamics of e-h pair radiative recombination in our system, a wide pulse with a full-pulse width around $300$\,ps is required, as displayed in Fig.\,\ref{f14}. From the upper-middle panel of
this figure, we find that $\beta_0(\omega_{\rm sp})$ starts with a resonant dip due to a strong photon dressing effect, then shifts to a sole peak at $\Delta\hbar\omega_{\rm sp}=0$ as $\tau_0\geq 400$\,ps
where a steady state is almost reached in the linear-response regime. Accordingly, the level populations $n_{\rm 1,e}$ and $n_{\rm 2,e}$ in the lower two panels show a transition from an initial non-resonant behavior
to a final resonant behavior. This is accompanied by dramatically reduced level populations due to the start of a radiative recombination process for photo-excited e-h pairs.
\medskip

\begin{figure}[p]
\centering
\epsfig{file=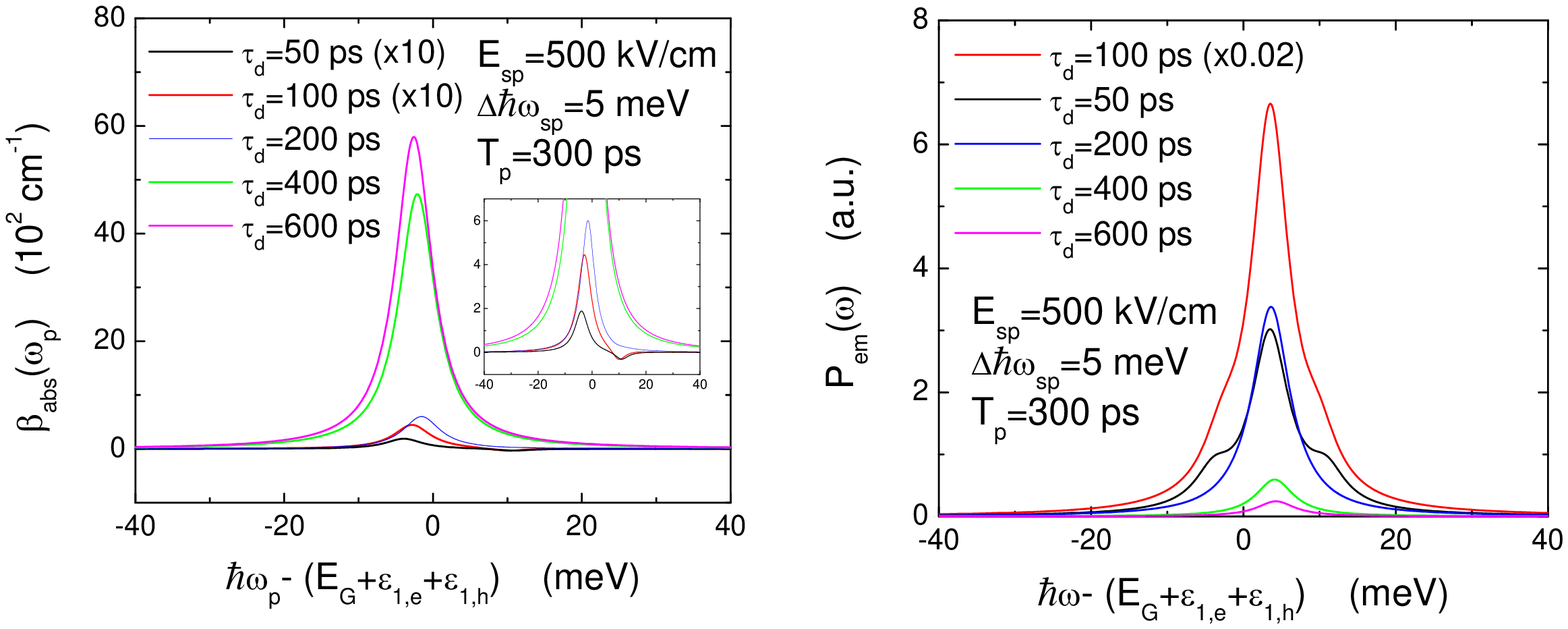,width=0.9\textwidth}
\caption{\label{f15}
(Color online) $\beta_{\rm abs}(\omega_{\rm p})$ (left panel) and $P_{\rm em}(\omega)$ (right panel)
as a function of $\hbar\omega_{\rm p}-(E_{\rm G}+\varepsilon_{\rm 1,e}+\varepsilon_{\rm 1,h})$ and $\hbar\omega-(E_{\rm G}+\varepsilon_{\rm 1,e}+\varepsilon_{\rm 1,h})$, respectively,
at different delayed times $\tau_d$ for a Gaussian-shape laser pulse with pulse width $T_p=300$\,ps.
Here, $E_{\rm sp}=500$\,kV/cm and $\Delta\hbar\omega_{\rm sp}=5$\,meV are assumed, and the other parameters are the same as those in Fig.\,\ref{f3}.
The inset in the upper-left panel shows an enlarged view for the gain and absorption peaks for small delayed times.
The labels $\times 10$ for $\tau_d=50$ and $100$\,ps in the left panel, as well as the label $\times 0.02$ in the right panel, indicate that the results are multiplied by a factor of $10$ and a factor of
$0.02$, respectively.}
\end{figure}

Recombination dynamics for e-h plasmas can also be demonstrated clearly by the time-delayed probe-field absorption as well as by the time-resolved spontaneous emission, as shown in Fig.\,\ref{f15}.
As presented in the left panel of this figure, we find that the initial weak absorption and gain peaks (see the inset) in $\beta_{\rm abs}(\omega_{\rm p})$
occur at $\tau_d=200$\,ps and are replaced by a strong single absorption peak due to a suppressed photon dressing effect and
phase-space blocking. On the other hand, from the right panel of the same figure, we see that the initial central peak in $P_{\rm em}(\omega)$ is increased very rapidly due to accumulation
of photo-excited e-h pairs and accompanied by the reduction of two side peaks resulting from the weakened Rabi oscillations. Importantly, the very-strong central peak in $P_{\rm em}(\omega)$
is significantly reduced at $\tau_d=200$\,ps, indicating the start of a radiative-recombination process for photo-excited e-h plasmas.
This recombination process is continuously enhanced with the increasing delay time $\tau_d$ and suppresses the central peak in $P_{\rm em}(\omega)$ after
$\tau_d\geq 400$\,ps due to draining out the photo-generated electrons and holes at the same time.

\section{Conclusions and Remarks}
\label{sec4}

In conclusion, we have demonstrated the possibility of using a SPP field to control the optical gain and absorption of another passing light beam due to their strong
nonlinear field coupling mediated by electrons in the quantum dot. We have also predicted the coherent conversion of a surface-plasmon-field photon to a spontaneously-emitted free-space photon,
which is simultaneously accompanied by another pair of blue- and red-shifted photons.
\medskip

Although we studied only the coupling of a SPP field to a single quantum dot in this paper for the simplest case, our formalism can be generalized easily to
include many quantum dots. The numerically-demonstrated unique control of the effective photon-photon coupling by the quantum dot can be used for constructing an optical transistor,
where the `gate' photon controls the intensity of its `source' light beam.
These optical transistors are very useful for speeding up and improving the performance of fiber-optic communication networks, as well as for constructing quantum information and developing optical digital computers.
\medskip

Furthermore, instead of a resonant coupling to the lowest pair of electron-hole energy levels, we may select the surface-plasmon frequency for resonant coupling to the higher pair of electron-hole levels.
In such a case, the optical pumping from the intense surface-plasmon near-field could create a population inversion with respect to the ground pair of electron-hole levels by emitting thermal phonons,
leading to a possible lasing action if the optical gain can overcome the metal loss for the surface plasmons. Such a surface-plasmon quantum-dot laser would have a beam size as small as a few nanometers beyond the optical diffraction limit, and it is expected to be very useful for spatially-selective illumination of individual molecules or neuron cells in low-temperature photo-excited chemical reactions or optogenetics and neuroscience.

\begin{acknowledgments}
DH would like to thank the support from the Air Force Office of Scientific Research (AFOSR).
\end{acknowledgments}

\appendix

\section{Electronic States of a Quantum Dot}

We have employed a box-type potential with hard walls for a quantum dot, which is given by

\begin{equation}
V({\bf r})=\left\{\begin{array}{ll}
0\ , & \mbox{$0\leq x_i\leq L_i$ for $i=1,\,2,\,3$}\\
\infty\ , & \mbox{others}
\end{array}\right.\ ,
\label{a1}
\end{equation}
where the position vector ${\bf r}=(x_1,x_2,x_3)$, $L_1$, $L_2$ and $L_3$ are the widths of the potential in the $x_1$, $x_2$ and $x_3$ directions, respectively. The Schr\"odinger equation for a single electron or hole in a quantum dot is written as

\begin{equation}
-\frac{\hbar^2}{2m^\ast}\left[\frac{\partial^2}{\partial x_1^2}+\frac{\partial^2}{\partial x_2^2}+\frac{\partial^2}{\partial x_3^2}+V({\bf r})\right]\psi({\bf r})=\varepsilon\,\psi({\bf r})\ ,
\label{a2}
\end{equation}
where the effective mass $m^\ast$ is $m^\ast_{\rm e}$ for electrons or $m^\ast_{\rm h}$ for holes. The eigenstate wave-function associated with Eq.\,(\ref{a2}) is found to be

\begin{equation}
\psi_{n_1,n_2,n_3}({\bf r})=\sqrt{\frac{2}{L_1}}\,\sin\left[\left(\frac{n_1\pi}{L_1}\right)x_1\right]\sqrt{\frac{2}{L_2}}\,\sin\left[\left(\frac{n_2\pi}{L_2}\right)x_2\right]
\sqrt{\frac{2}{L_3}}\,\sin\left[\left(\frac{n_3\pi}{L_3}\right)x_3\right]\ ,
\label{a3}
\end{equation}
which is same for both electrons and holes, and the eigenstate energy associated with Eq.\,(\ref{a2}) is

\begin{equation}
\varepsilon_{n_1,n_2,n_3}=\frac{\hbar^2}{2m^\ast}\left[\left(\frac{n_1\pi}{L_1}\right)^2+\left(\frac{n_2\pi}{L_2}\right)^2+\left(\frac{n_3\pi}{L_3}\right)^2\right]\ ,
\label{a4}
\end{equation}
where the quantum numbers $n_1,\,n_2,\,n_3=1,\,2,\,\cdots$.
\medskip

By using the calculated bare energy levels in Eq.\,(\ref{a4}), the dressed electron ($\lambda^{\rm e}_\alpha$) and hole ($\lambda^{\rm h}_\alpha$) energy levels under the rotating wave approximation take the form of\,\cite{r7}

\begin{equation}
\lambda^{\rm e}_\alpha(\omega\vert t)=\lambda^{\rm h}_\alpha(\omega\vert t)=
\left\{\begin{array}{cc}
\frac{1}{2}\left(\hbar\omega+\sqrt{[{\cal E}_{\rm G}(T)+\varepsilon_\alpha^{\rm e}+\varepsilon_\alpha^{\rm h}-\hbar\omega]^2+4|{\cal M}^{\rm eh}_{\alpha,\alpha}(t)|^2}\,\right) &
\\
\mbox{if $\hbar\omega\leq {\cal E}_{\rm G}(T)+\varepsilon_\alpha^{\rm e}+\varepsilon_\alpha^{\rm h}$}\\
\\
\frac{1}{2}\left(\hbar\omega-\sqrt{[{\cal E}_{\rm G}(T)+\varepsilon_\alpha^{\rm e}+\varepsilon_\alpha^{\rm h}-\hbar\omega]^2+4|{\cal M}^{\rm eh}_{\alpha,\alpha}(t)|^2}\,\right) &
\\
\mbox{if $\hbar\omega\geq {\cal E}_{\rm G}(T)+\varepsilon_\alpha^{\rm e}+\varepsilon_\alpha^{\rm h}$}
\end{array}\right.\ ,
\label{a5}
\end{equation}
where the composite index $\alpha=\{n_1,\,n_2,\,n_3\}$. Moreover, we get the energy levels of dressed electrons $\overline{\varepsilon}^{\rm e}_{\alpha}(\omega\vert t)=\lambda^{\rm e}_\alpha(\omega\vert t)+(\varepsilon^{\rm e}_\alpha-\varepsilon^{\rm h}_\alpha)/2$ and $\overline{\varepsilon}^{\rm e}_{\ell}(\omega\vert t)=\varepsilon_{\ell}^{\rm e}+{\cal E}_{\rm G}(T)/2$ for $\ell\neq\alpha$. Similarly, we obtain the energy levels of dressed holes $\overline{\varepsilon}^{\rm h}_{\alpha}(\omega\vert t)=\lambda^{\rm h}_\alpha(\omega\vert t)+(\varepsilon^{\rm h}_\alpha-\varepsilon^{\rm e}_\alpha)/2$
and $\overline{\varepsilon}^{\rm e}_j(\omega\vert t)=\varepsilon_j^{\rm h}+{\cal E}_{\rm G}(T)/2$ for $j\neq\alpha$.
\medskip

Based on the calculated wave-functions in Eq.\,(\ref{a3}), the form factors introduced in Eqs.\,(\ref{e11}) and (\ref{e12}) can be obtained from

\begin{equation}
{\cal F}^{\rm e}_{n_1,n_2,n_3;\,n_1^\prime,n_2^\prime,n_3^\prime}({\bf q})={\cal F}^{\rm h}_{n_1,n_2,n_3;\,n_1^\prime,n_2^\prime,n_3^\prime}({\bf q})={\cal Q}^{(1)}_{n_1,n_1^\prime}(q_1)\,{\cal Q}^{(2)}_{n_2,n_2^\prime}(q_2)\,{\cal Q}^{(3)}_{n_3,n_3^\prime}(q_3)\ ,
\label{a6}
\end{equation}
where the wave vector ${\bf q}=(q_1,q_2,q_3)$ and we have defined the following notation for $j=1,\,2,\,3$

\begin{equation}
{\cal Q}^j_{n_j,n_j^\prime}(q_j)=\left(\frac{2}{L_j}\right)\int\limits_0^{L_j} dx_j\,{\rm e}^{iq_jx_j}\,\sin\left[\left(\frac{n_j\pi}{L_j}\right)x_j\right]\sin\left[\left(\frac{n^\prime_j\pi}{L_j}\right)x_j\right]\ .
\label{a7}
\end{equation}
Moreover,the overlap of the electron and hole wave-functions in this model can be easily calculated as

\begin{equation}
\int d^3{\bf r}\,\psi^{\rm e}_{n_1,n_2,n_3}({\bf r})\,\psi^{\rm h}_{n^\prime_1,n^\prime_2,n^\prime_3}({\bf r})=\delta_{n_1,n^\prime_1}\,\delta_{n_2,n^\prime_2}\,\delta_{n_3,n^\prime_3}\ .
\label{a8}
\end{equation}
The interband dipole moment ${\bf d}_{\rm c,v}=d_{\rm c,v}\,\hat{\bf e}_{\rm d}$ at the isotropic $\Gamma$-point, which is defined in Eq.\,(\ref{e32}), can be calculated according to the Kane approximation\,\cite{r14,r15}

\begin{equation}
d_{\rm c,v}=\sqrt{\frac{e^2\hbar^2}{2m_0\,{\cal E}_{\rm
G}(T)}\left(\frac{m_0}{m^\ast_{\rm e}}-1\right)}\ . \label{a9}
\end{equation}
Furthermore, the direction of the dipole moment $\hat{\bf e}_{\rm d}$ is determined by the quantum-dot energy levels in resonance with the photon energy $\hbar\omega$.

\newpage

\end{document}